\documentclass[sigconf,screen]{acmart}

\usepackage{todos}
\usepackage{balance}
\usepackage{booktabs}
\usepackage{multirow}
\usepackage{booktabs}
\usepackage{graphicx}
\usepackage{epsfig}
\usepackage{grffile}%check the file true extension
\usepackage{algorithm}
\usepackage{subfig}
\usepackage{listings}
\usepackage{hyperref}
\usepackage{paralist}
\usepackage{xspace}
\usepackage[super]{nth}
\usepackage[T1]{fontenc}
\usepackage{setspace}
\hypersetup{
    colorlinks=true,
    linkcolor=black,
    citecolor=black,
    filecolor=black,
    urlcolor=black,
}
\usepackage{xcolor}
\usepackage{tabularx}
\usepackage{csquotes}
\usepackage{dirtytalk}
\usepackage[framemethod=default]{mdframed}
\usepackage{showexpl}
\usepackage{epigraph}
\usepackage{microtype}

\mdfsetup{skipabove=4pt,skipbelow=4pt}

\definecolor{pblue}{rgb}{0.13,0.13,1}
\definecolor{darkcyan}{rgb}{0.0, 0.55, 0.55}
\definecolor{pgreen}{rgb}{0.37, 0.62, 0.63}
\definecolor{pred}{rgb}{0.9,0,0}
\definecolor{pgrey}{rgb}{0.46,0.45,0.48}

\makeatletter
\newcommand\BeraMonottfamily{%
  \def\fvm@Scale{0.85}% scales the font down
  \fontfamily{fvm}\selectfont% selects the Bera Mono font
}
\makeatother
\lstdefinestyle{interfaces}{
	float=tp,
	floatplacement=tbp,
	abovecaptionskip=0pt
}

\DeclareMathOperator*{\argmax}{arg\,max}
\newtheorem{mydef}{Definition}

\newcommand{\sect}[1]{\autoref{#1}\xspace}

\newcommand{\longcomment}[1]{}
\newcommand{\op}[1]{\texttt{#1}\xspace}

\renewcommand{\P}{{\cal P}}

\newcommand{\stitle}[1]{\noindent \textbf{#1}\hspace{2ex}}

\newcommand{\acr}[1]{\textsc{\lowercase{#1}}}

\newcommand{\systemName}{\acr{ARTEMIS}}

\newcommand{\eg}{\hbox{\emph{e.g.,}}\xspace}

\begin{document}

\copyrightyear{2018} 
\acmYear{2018} 
\setcopyright{acmlicensed}
\acmConference[ESEC/FSE '18]{Proceedings of the 26th ACM Joint European Software Engineering Conference and Symposium on the Foundations of Software Engineering}{November 4--9, 2018}{Lake Buena Vista, FL, USA}
\acmBooktitle{Proceedings of the 26th ACM Joint European Software Engineering Conference and Symposium on the Foundations of Software Engineering (ESEC/FSE '18), November 4--9, 2018, Lake Buena Vista, FL, USA}
\acmPrice{15.00}
\acmDOI{10.1145/3236024.3236043}
\acmISBN{978-1-4503-5573-5/18/11}

\title{Darwinian Data Structure Selection}

% \numberofauthors{5}
 \author{Michail Basios, Lingbo Li, Fan Wu, Leslie Kanthan, Earl T. Barr}
 \affiliation{%
   \institution{University College London, UK}
%   \country{UK}
   % \streetaddress{2 Malet Pl}
   % \city{Camden Town}
   % \state{London}
   % \postcode{WC1E 7JG}
 }
 \email{{michail.basios,lingbo.li,fan.wu,l.kanthan,e.barr}@cs.ucl.ac.uk}

\begin{abstract}

Data structure selection and tuning is laborious but can vastly improve an
  application's performance and memory footprint.  Some data structures share a
  common interface and enjoy multiple implementations.  We call them Darwinian
  Data Structures (DDS), since we can subject their implementations to survival of the fittest.  We introduce \systemName{} a multi-objective,
  cloud-based search-based optimisation framework that automatically finds
  optimal, tuned DDS modulo a test suite, then changes an application to use that
  DDS.  \systemName{} achieves substantial performance improvements for
  \emph{every} project in $5$ Java projects from DaCapo benchmark, $8$
  popular projects and $30$ uniformly sampled projects from GitHub.  For
  execution time, CPU usage, and memory consumption, \systemName{} finds at
  least one solution  that improves \emph{all} measures for $86\%$ ($37/43$)
  of the projects.  The median improvement across the  best solutions is $4.8\%$, $10.1\%$, $5.1\%$ for runtime, memory and CPU usage.
%  for execution time, $10.1\%$ for memory consumption and $5.1\%$ for CPU
%  usage.

These aggregate results understate \systemName{}'s potential impact.  Some of
  the benchmarks it improves are libraries or utility functions.  Two examples
  are \lstinline+gson+, a ubiquitous Java serialization framework, and
  \lstinline+xalan+, Apache's XML transformation tool.
  \systemName{} improves \lstinline+gson+ by $16.5$\%, $1\%$ and $2.2\%$ for memory, runtime, and CPU;
  \systemName{} improves \lstinline+xalan+'s memory consumption by $23.5$\%.
  \emph{Every} client of these projects will benefit from these performance improvements.

\end{abstract}

% <concept_id>10011007.10011074.10011111.10011113</concept_id>
% <concept_desc>Software and its engineering~Software evolution</concept_desc>
% <concept_id>10011007.10011074.10011784</concept_id>
% <concept_desc>Software and its engineering~Search-based software engineering</concept_desc>
% \ccsdesc[500]{Software and its engineering~Software evolution}
% \ccsdesc[500]{Software and its engineering~Search-based software engineering}




\settopmatter{printfolios=true}
\maketitle

\epigraph{
% \vspace{-1em}
\textit{``Programmers waste enormous amounts of time thinking about, or worrying about, the speed of noncritical parts of their programs, and these attempts at efficiency actually have a strong negative impact when debugging and maintenance are considered. We should forget about small efficiencies, say about 97\% of the time: premature optimization is the root of all evil. Yet we should not pass up our opportunities in that critical 3\%.''}}{--- Donald E. Knuth~\cite{Knuth:1974:SPG:356635.356640}
\vspace{-2em}
}

\section{Introduction}
\label{sec:introductions}

Under the immense time pressures of industrial software development, developers are heeding one part of Knuth's advice:  they are avoiding premature optimisation.  
Indeed, developers appear to be avoiding optimisation altogether and neglecting the ``critical $3\%$".  
When selecting data structures from libraries, in particular, they tend to rely on defaults and neglect potential optimisations that alternative implementations or tuning parameters can offer. 
This, despite the impact that data structure selection and tuning can have on application performance and defects. 
Consider three examples. Selecting an implementation that creates unnecessary temporary objects for the program's workload~\cite{xu2009go}. Selecting a combination of Scala data structures that scaled better,  reducing execution time from $45$ to $1.5$ minutes~\cite{story2}. Avoiding the use of poor implementation, such as those in the Oracle bug database that leak memory~\cite{xu2008precise}.

%This, despite the impact that data structure selection and tuning can have on application performance and defects. 
%For performance, examples include the selection of an implementation that created unnecessary temporary objects for the program's workload~\cite{xu2009go} or selecting a combination of Scala data structures that scaled better, reducing execution time from $45$ to $1.5$ minutes~\cite{story2}; memory leak bugs exemplify data structure triggered defects, such as those in the Oracle Java bug database caused by poor implementations that retained references to unused data entries~\cite{xu2008precise}. 

Optimisation is time-consuming, especially on large code bases. 
It is also brittle. An optimisation for one version of a program can break or become a de-optimisation in the next release. 
Another reason developers may avoid optimisation are development fads that focus on fast solutions, like ``Premature Optimisation is the horror of all Evil" and ``Hack until it works"~\cite{story3}.  
In short, optimisation is expensive and its benefits unclear for many projects.  
Developers need automated help.

Data structures are a particularly attractive optimisation target because they
have a well-defined interface; many are tunable; and different implementations
of a data structure usually represent a particular trade-off between time and
storage, making some operations faster but more space-consuming or slower but
more space-efficient.  For instance, an ordered list makes retrieving the
entire dataset in sorted order fast, but inserting new elements slow, whilst a
hash table allows for quick insertions and retrievals of specific items, but
listing the entire set in order is slow.  We introduce 
\emph{Darwinian data structures}, distinct data structures that are interchangeable 
because they share an abstract data type and can be tuned.
The Darwinian data structure
optimisation problem is the problem of finding an optimal 
implementation and tuning for a Darwinian data structure used in an input program.  

%We aim to help developers perform optimisation cheaply, focusing solving the data structure optimisation problem.  
%We present \systemName{}, a cloud-based optimisation framework that identifies \emph{Darwinian data structures} and, given a test suite, \emph{automatically} searches for optimal combinations of implementations and parameters for them.  
%\systemName{} is language-agnostic; we have instantiated it for Java and C++, and present optimisation results for both languages (\autoref{sec:evaluation}). 
%\systemName{}' search is multi-objective, seeking to simultaneously improve a program's execution time, memory usage, and CPU usage while passing all the test suites. 
%\systemName{} scales to large code bases by focusing genetic improvement search in the area with the most valid solutions (\autoref{subsection:optimiser}). 
%\systemName{} is the first technique to apply multi-objective optimisation to the Darwinian data structure selection and tuning problem.

We aim to help developers perform optimisation cheaply, focusing solving the data structure optimisation problem.  
We present \systemName{}, a cloud-based optimisation framework that identifies \emph{Darwinian data structures} and, given a test suite, \emph{automatically} searches for optimal combinations of implementations and parameters for them.  
\systemName{} is language-agnostic; we have instantiated it for Java and C++, and present optimisation results for both languages (\autoref{sec:evaluation}). 
\systemName{}' search is multi-objective, seeking to simultaneously improve a program's execution time, memory usage, and CPU usage while passing all the test suites. 
\systemName{} scales to large code bases because is uses a Genetic algorithm on those regions of its search space with the most solutions (\autoref{subsection:optimiser}).
\systemName{} is the first technique to apply multi-objective optimisation to the Darwinian data structure selection and tuning problem.

% \todo[Mike]{Earl comment, A todo for the intro please brag about how Artemis
% makes economical small optimizations optimizations that wont pay for
% themselves in developer time and that, in the search for these small
% optimizations, it can also find unexpected big ones}

\systemName{} promises to change the economics of data structure optimisation.
Given a set of Darwinian data structures, \systemName{} can search for optimal
solutions in the background on the cloud, freeing developers to focus on new
features. \systemName{} makes economical small optimizations, such as a few 
percent, that would not pay for the developer time spent realizing them.
And sometimes, of course, \systemName{}, by virtue of being used, will find
unexpectedly big performance gains.

\systemName{} is a source-to-source transformer.
When \systemName{} finds a solution, the program variants it produces 
differ from the original program only at constructor calls and relevant type
annotations.  Thus, \systemName{}' variants are amenable, by design, to
programmer inspection and do not increase technical
debt~\cite{brown2010managing}.  To ease inspection, \systemName{} generates a
diff for each changes it makes.  Developers can inspect these diffs and decide
which to keep and which to reject.

We report results on $8$ popular diverse GitHub projects, on
DaCapo benchmark which was constructed to be representative, and a corpus of
$30$ GitHub projects, filtered to meet \systemName{}'s constraints and sampled
uniformly at random.  In this study, \systemName{} achieves substantial
performance improvements for all $43$ projects in its corpus.  In terms of
execution time, CPU usage, and memory consumption, \systemName{} finds at least
one solution for $37$  out of $43$ projects that improves \emph{all} measures.
Across all produced optimal solutions, the median improvement for execution
time is $4.8\%$, memory consumption $10.1\%$ and CPU usage $5.1\%$.  This
result is for various corpora, but it is highly likely to generalise to
arbitrary programs because of the care we took to build a representative 
corpus (\autoref{subsec:corpus}).

These aggregate results understate \systemName{}'s potential impact.  Some of
our benchmarks are libraries or utilities.  All of their clients will enjoy any
improvements \systemName{} finds for them.  Three examples are the Apache
project's powerful XSLT processor \op{xalan}, \op{Google-http-java-client}, the unbiquitious Java library for
accessing web resources, and  Google's in-memory
file system \op{Jimfs}.  \autoref{sec:evaluation} shows that \systemName{}
improved \op{xalan}'s memory consumption by $23.5\%$, while leaving its
execution time unchanged;  \systemName{} improved
\op{Google-http-java-client}'s execution time by $46$\% and its CPU usage by
$39.6$\%; finally, \systemName{}  improved \op{Jimfs}'s execution time by
$14.2\%$ and its CPU usage by $10.7$\%, while leaving its memory consumption
unchanged.

Our principal contributions follow:

\begin{itemize}
    \item We formalise the Darwinian data structure selection and optimisation problem \textbf{DS\textsuperscript{2}} (\sect{sec:problem_formulation}).
    \item We implement \systemName{}, a multi-language optimisation framework that automatically discovers and optimises sub-optimal data structures and their parameters.

    \item We show the generalizability and effectiveness of \systemName{} by
      conducting a large empirical study on a corpus comprising $8$ popular
      GitHub project, $5$ projects from the DaCapo benchmark, and $30$ Github
      projects, filtered then sampled uniformly.  For all $43$ subjects,
      \systemName{}  find variants that outperforms the original
      for all three objectives.  

      %In extreme cases, \systemName{} discovered $46\%$ improvement on
      %execution time, $44.9\%$ improvement on memory consumption, and $49.7\%$
      %improvement on CPU usage.

    \item We provide \systemName{} as a service, along with its code and 
      evaluation artifacts at \url{http://darwinianoptimiser.com}.
    
\end{itemize}

\section{Motivating example}
\label{sec:motivating_example}

%% The first several paragraphs have been moved to the introduction section.

\autoref{fig:code1} contains a code snippet from \op{google\char`-http\char`-java\char`-client}\footnote{\url{https://github.com/google/google-http-java-client}}, a popular  Java library for accessing efficiently resources on the web. In the \autoref{fig:code1}, \op{getAsList} packages HTTP headers and is invoked frequently from other methods because they use it every time they construct an HTTP request. Thus, its functionality is important for the performance of \op{google\char`-http\char`-java\char`-client}.

%{\footnotesize
%\begin{minipage}{0.8\linewidth}
%\captionsetup{font=small}
%\centering 
%\begin{lstlisting}[language=Java,numbers=left, caption=A function from http-java-client., label=fig:code1, escapechar=|][float,floatplacement=H]
%<T> List<T> getAsList(T value) {
%  if (value == null)
%    return null;
%  List<T> result = new ArrayList<T>(); |\label{fig:code1:line4}|
%  result.add(value);
%  return result;
%}   
%\end{lstlisting}
%\setlength{\belowcaptionskip}{5pt}
%\end{minipage}
%}

\begin{lstlisting}[xleftmargin=3.5ex,label=fig:code1,style=interfaces,escapechar=|,language=Java,numbers=left,caption={A function from http-java-client.}]
<T> List<T> getAsList(T value) {
	if (value == null)
		return null;
	List<T> result = new ArrayList<T>(); |\label{fig:code1:line4}|
	result.add(value);
	return result;
}   
\end{lstlisting}

\autoref{fig:code1} uses \lstinline{ArrayList} to instantiate the \op{result}
variable.  However, other \lstinline{List} implementations share the same
functionality but different non-functional properties.  Thus, replacing
\lstinline{ArrayList} with other \lstinline{List} implementations may affect
the performance of the program.  Considering the variant created when replacing
\lstinline{ArrayList} (\autoref{fig:code1}, line~\ref{fig:code1:line4}) with \lstinline{LinkedList},
when we compare it with the original program against the same test set for $30$
runs~(\sect{sec:implementation}), we see that the \op{google-http-java-client} achieves a median $46\%$,
with 95\% Confidence Interval [$45.6\%$, $46.3\%$] improvement in execution
time (\autoref{sec:evaluation}).

\systemName{}, our optimization framework, automatically discovers
underperforming data structures and replaces them with better choices using
search-based techniques~(\autoref{subsection:optimiser}).  First, it 
\emph{automatically} creates a store of data structures from the language's Collection
\acr{API} library (\autoref{subsec:store}). Then, \systemName{} traverses 
the program's \acr{AST} to identify which of those data structures are used 
and exposes them as parameters to the \systemName{}'s \acr{optimizer}~(\autoref{subsection:optimiser}) by transforming 
line~\ref{fig:code1:line4} into 

{\small
\begin{minipage}{.9\linewidth}
\centering
\begin{lstlisting}[language=Java]
List<T> result = new D<T>();
\end{lstlisting}
\setlength{\belowcaptionskip}{5pt}
\end{minipage}
}

\noindent where $D$ is the tag that refers to the exposed parameter associated with the
defined data structure type (\sect{sec:implementation}).

% For example, a data structure \textit{A} that consumes less memory but has worse execution time than a data structure \textit{B} does not necessarily mean that will provide a program that is faster. Huge memory consumption may trigger more frequently the Garbage Collection and thus making it slower.

% \item
% \textbf{Data Structure Parameter Tuning:}
%\autoref{fig:code1} does not specify the initial capacity size of the \lstinline{ArrayList}, so the default size $10$ was used. 
%Thus, when the instantiated \lstinline{List} object only contains one item, the default capacity  can result in memory bloat. 
%Tuning the initial capacity size of the \lstinline{ArrayList} enables to control the amount of memory pre-allocated. 
%However, an inappropriate large value of $S$ may waste memory, while an overly small value may force the program to allocate new memory frequently during the execution, slowing the execution. 
%Therefore, an appropriate value must be chosen to simultaneously balance memory and performance.

\autoref{fig:code1} does not specify the initial capacity size of the \lstinline{ArrayList}, so the default size $10$ was used. 
If the instantiated \lstinline{List} object contains less than $10$ items, the default capacity can result in memory bloat. 
If the \lstinline{List} object contains more than $10$ items, the default capacity can slow the execution time; more memory realllocation operations will happen.
Therefore, an appropriate value must be chosen to find a good tradeoff between memory and execution time.

%The choice of the defaul size value of the list affects both memory consumption and execution time of the program. \lstinline{ArrayList}  is a dynamically resizing array data structure, implemented as an array with an initial (default) fixed size that is extended to a double size when it gets filled up. Reallocating memory is a costly operation and it
% When this gets filled up, the array will be extended to a double sized one. This operation is costly, so you want as few as possible.

\systemName{} automatically exposes such arguments as tunable parameters, then adjusts them to improve the runtime performance of the program. 
For instance, \systemName{} changes line~\ref{fig:code1:line4} to the code below:

{\small
\begin{minipage}{.9\linewidth}
\centering
\begin{lstlisting}[language=Java]
List<T> l = new ArrayList<>(S);
\end{lstlisting}
\setlength{\belowcaptionskip}{5pt}
\end{minipage}
}

\noindent where
$S$ refers to the exposed parameter associated with the amount of pre-allocated memory.

%Our tool\footnote[3]{\url{https://omitted to meet double-blind review requirements}} can solve automatically this bloat problem. 
% The following command shows an example of using \systemName{} :

% {\small
% \begin{minipage}{.9\linewidth}
% \centering
% \begin{lstlisting}[language=Java]
% ./artemis google-http-java-client-src output-src
% \end{lstlisting}
% % \setlength{\belowcaptionskip}{2pt}
% \end{minipage}
% }

%The result of optimising the \op{google-http-java-client} by merely optimising this code snippet shows that the proposed tool can automatically boost $9.9\%$ execution time and reduce $47.1\%$ memory usage when executes all test suits.

% Also, the user should note that choosing a version of $D$ may affect the existence of parameter $S$; e.g. a $LinkedList$ does not accept initial $S$ parameter.

% Next, we will show a graph with performance difference between those three versions of the code. We should show details about execution time, memory consumption and CPU usage. We should also point out the difficulty of finding the correct data structure and how that decision may be wrong if the importance of the metrics change. For example, if code needs to run in a device with less memory than the memory optimisation may be more significant than the execution time. 

\section{Darwinian Data Structure Selection and Tuning}
\label{sec:problem_formulation}

\def\Replaceable{\phi}

This section defines the Darwinian data structure and parameter optimisation
problem we solve in this paper.

\begin{mydef}
[Abstract Data Type] An \emph{Abstract Data Type} (ADT) is class of objects
whose logical behavior is defined by a set of values and a set of
operations~\cite{dale1996abstract}.
\end{mydef}

A \emph{data structure} concretely implements an ADT.  For the corpus of
programs $C$ and the ADT $a$, the data structure extraction function
$\mathit{dse}(a,C)$ returns all data structures that implement $a$ in $C$.
This function is integral to the definition that follows.

\begin{mydef}[Darwinian Data Structure] 
  \label{def:dds}
When $\exists d_0,d_1 \in \mathit{dse}(a,C) \wedge d_0 \ne d_1 \wedge d_0$ and
  $d_1$  are observationally equivalent modulo $a$, $d_0$ and $d_1$ are
  \emph{Darwinian data structures}.
\end{mydef}

In words, Darwinian data structures are $darwinian$ in the sense that they can be replaced to produce program mutants whose fitness we can evaluate. The ADT $a$ has Darwinian data structures when it has more than one
data structure that are equivalent over the operations the ADT defines.  In Java,
\lstinline{List} is an ADT and \lstinline{ArrayList}, which implements it, is a
data structure.  \lstinline{LinkedList} also implements \lstinline{List}, so
both \lstinline{ArrayList} and \lstinline{LinkedList} are Darwinian.  For the
ADT $a$ and the corpus $C$, Darwinian data structures are interchangeable.
Thus, we can search the variants of $P \in C$ formed by substituting one
Darwinian data structure for another to improve $P$'s nonfunctional properties,
like execution time, memory consumption or CPU usage.

Just as we needed a function to extract an ADT's data structures from a corpus
for \hyperref[def:dds]{Definition 2}, we need a function that returns the ADT that a data
structure implements: when $d = \mathit{dse}(a,C)$, let $\mathit{adte}(d,C) =
a$.  Let $\Gamma_D$ bind fully scope-qualified declarations of names to
Darwinian data structures in $C$.  We use $\Gamma_D$ when creating variants of
a program via substitution.  We are interested not just searching the space of Darwinian data structures, but also tuning them via their constructor
parameters.  To this end, we assume without loss of generality that $a$ defines a
single constructor $c$ and we let $n.c(x)$ denote calling identifier $n$'s
constructor $c$ with parameters $x : \tau$.

To create a variant of $P \in C$ that differs from $P$ only in its $k$ bindings
of names to Darwinian data structures or their constructor initialization 
parameter, we define 

$\phi(P,(n,d_i)^k,d_j^k,x_j) =$ \\
\vspace*{-4mm}
\begin{align*}
  \text{\qquad}
  \begin{cases}
    P[(n.c(x_i))^k / (n.c(x_j))^k],
       \text{ if } \exists d_i, d_j \text{ s.t. } \mathit{adte}(d_i) \ne \mathit{adte}(d_j) \\
    P[(n,d_i)^k / (n,d_j)^k][(n.c(x_i))^k / (n.c(x_j))^k], \text{ otherwise}
  \end{cases}
\end{align*}

\begin{mydef}[Darwinian Data Structure Selection and Tuning]
For the real-valued fitness function $f$ over the corpus $C$,
\emph{the Darwinian data structure and tuning problem} is 
\begin{align*}
  \argmax\limits_{(n_i,d_i)^k\in\Gamma_D^k,d_j^k\in\mathit{adte}(d_i,C)^k,x_j\in\tau} f(\phi(P,(n_i,d_i),d_j,x_j))
\end{align*}
\end{mydef}

%\textit{\begin{equation*}
%	\prod_{a \in P}^{} |I(a)|^{dse(a,P)} * | dom(a.c)|^{arity(a.c)}\text{, where d.c is d's constructor.}
%\end{equation*}}

This vector-based definition simultaneously considers all possible rebinding of
names to Darwinian data structures in $P$; it is also cumbersome, compared to
its point-substitution analogue.  We could not, however, simply present a
scalar definition and then quantify over all potential DDSS substitutions, as
doing so would not surface synergies and antagonisms among the substitutions.

\section{Artemis}
\label{sec:implementation}

\begin{figure*}[!tb]
  \centering
  \includegraphics[width=0.99\textwidth]{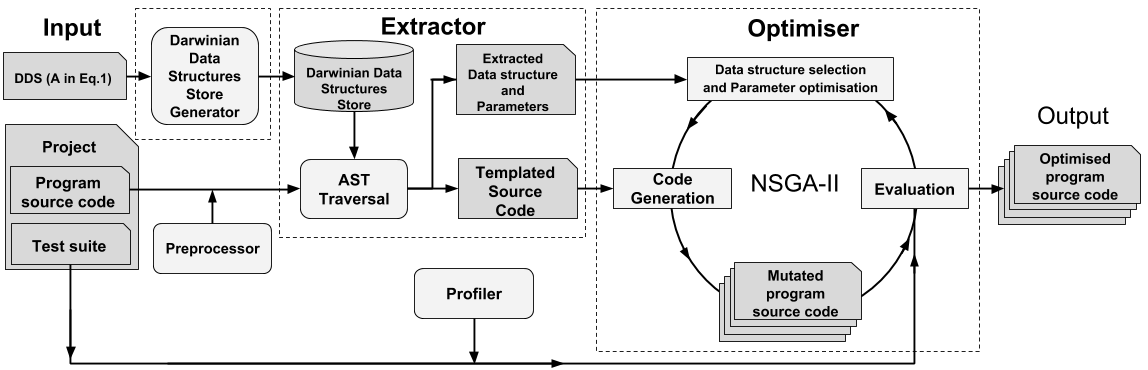}  
  \caption{System Architecture of \systemName{}.}
  \label{fig:system_architecture}
\end{figure*}

The \systemName{}'s optimisation framework solves the Darwinian Data Structure Selection problem.
\autoref{fig:system_architecture} illustrates the architecture with its three main components: the \acr{Darwinian Data Structures Store Generator} (\acr{DDSSG}), the \acr{Extractor}, and the \acr{Optimiser}. 
\systemName{} takes the language's Collection \acr{API} library, the user's application source code and a test suite as input to generate an optimised version of the code with a new combination of data structures. 
The \acr{DDSSG} builds a store that contains data structure transformation rules. The \acr{extractor} uses this store to discover potential data structure transformations and exposes them as tunable parameters to the \acr{optimiser} (see \sect{subsection:extractor}).
The \acr{optimiser} uses a multi-objective genetic search algorithm (NSGA-II~\cite{nsgaii}) to tune the parameters~\cite{fan2015, langdon2014improving,fan3,lingbo3,lingbo4} and to provide optimised solutions (see \sect{subsection:optimiser}). 
A regression test suite is used to maintain the correctness of the transformations and to evaluate the non-functional properties of interest. 
\systemName{} uses a built-in profiler that measures execution time, memory and CPU usage.

\systemName{} relies on testing to define a performance search space and to 
preserve semantics.  \systemName{} therefore can only be applied to programs
with a test suite.
Ideally, this test suite would comprise both a regression test suite with high
code coverage for maintaining the correctness of the program and a performance
test suite to simulate the real-life behaviour of the program and ensure that
all of the common features are covered~\cite{binder2000testing}. Even though
performance test suites are a more appropriate and logical choice for
evaluating the non-functional properties of the program, most real world
programs in GitHub do not provide such performance test suite. For this reason,
we use the regression test suites to evaluate the non-functional properties of
the GitHub projects of this study whenever a performance test suite is not
available.

\subsection{Darwinian Data Structure Store}
\label{subsec:store}

\begin{figure}[t]
  \centering
  \includegraphics[width=0.46\textwidth]{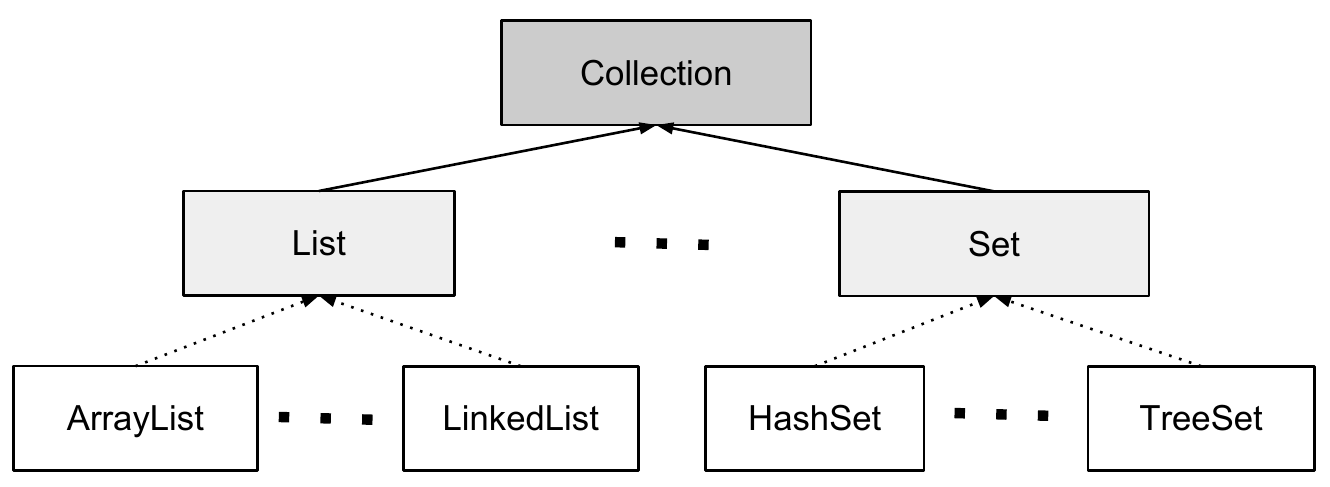}  
  \caption{DDS in the Java Collections API.}
  \label{fig:whitelist}
\end{figure}

\systemName{} needs a Darwinian data structure store (\acr{DDSS}) from which
to choose when creating variants. Let $A$ be a set of ADTs known to be Darwinian. A developer can augment this set; \autoref{fig:whitelist} shows those that \systemName{} knows by default. For our corpus $C$ of Java benchmarks 
augmented with JDK libraries over $A$, 
\begin{equation} \label{eq:1}
  DDSS = \bigcup_{a \in A} \mathit{dse}(a,C).
\end{equation}

To build the  default \acr{DDSS} for Java, \systemName{}
 extracts and traverses each project's class
hierarchy, similar to the one illustrated in \autoref{fig:whitelist}. This
hierarchy shows potential Darwinian data structures of a specific interface.
When this traversal finishes, \systemName{} extracts all the implementations of
a particular Darwinian data structure; \eg \lstinline{List}, \lstinline{ArrayList},
\lstinline{LinkedList}. \systemName{} considers these implementations mutually
replaceable. For Java, a default \acr{DDSS} is provided by \systemName{}, which the developer can edit. For other languages, the \acr{DDSS} can be provided manually by the user and this step can be skipped.  
The \acr{Optimiser}, described next, uses the store during its
search.

The developer can also extend the store with custom user-supplied implementations or with implementations from other third-party libraries such as Google Guava Collections\footnote{
\url{https://github.com/google/guava}}, fastutil\footnote{ \url{https://github.com/vigna/fastutil}} and Apache Commons Collections\footnote{ \url{https://github.com/apache/commons-collections}}.
% To take advantage of this option the user needs to add a parameter of the collection \acr{API} during the tool usage (\eg \lstinline{tool input output -api guava}) and \systemName{} will automatically import the necessary libraries and integrate those data structures as possible solutions.

\subsection{Discovering Darwinian Data Structures}
\label{subsection:extractor}

\begin{table}
  \centering
  \footnotesize
  \resizebox{0.5\textwidth}{!}{
  \begin{tabular}{rp{48mm}}
%   \begin{tabular}{@{}lllll@{}}
  \toprule
  Abstract Data Type & Implementation \\
  \midrule
List   & ArrayList, LinkedList  \\ 
Map   & HashMap, LinkedHashMap  \\ 
Set   & HashSet, LinkedHashSet  \\ 
Concurrent List & Vector, CopyOnWriteArrayList \\ 
Concurrent Deque & ConcurrentLinkedDeque, \newline LinkedBlockingDeque \\ 
Thread Safe Queue & ArrayBlockingQueue, SynchronousQueue, \newline LinkedBlockingQueue, DelayQueue, \newline ConcurrentLinkedQueue, LinkedTransferQueue \\
\hline 
  \end{tabular}
  }
  \caption{Data structure groups.}
  \label{table:1}
  \vspace{-3.5em}
\end{table}

The \acr{Extractor} takes as input the program $P$'s source code, identifies
Darwinian data structures in $P$ modulo its store (\autoref{subsec:store}),
and outputs a scope-qualified list of names of Darwinian data structures and
their constructor parameters (\emph{Extracted Data Structures and Parameters}
in \autoref{fig:system_architecture}).  For all
$a \in \text{DDSS}$, \acr{Extractor}'s output realises
$\mathit{dse}(a,P)$ (\autoref{sec:problem_formulation}). To mark potential substitions to the transformer, the \acr{Extractor} outputs a templated version of the code
which replaces the data structure with data structure type identifiers
(\emph{Templated Source Code} in \autoref{fig:system_architecture}).

To find \acr{darwinian} data structures, the \acr{Extractor} builds an Abstract Syntax Tree (\acr{AST}) from its input source code.
It then traverses the \acr{AST} to discover potential data structure transformations based on a store of data structures as shown in \autoref{table:1}.
For example, when an expression node of the \acr{AST} contains a \lstinline{LinkedList} expression, the \acr{extractor} marks this expression as a potential \acr{darwinian} data structure that can take values from the available \lstinline{List} implementations: \lstinline{LinkedList} or \lstinline{ArrayList}. 
The \acr{Extractor} maintains a copy of the \acr{AST}, referred to as the \acr{rewriter}, where it applies transformations, without changing the initial \acr{AST}. 
When the \acr{AST} transformation finishes, the \acr{Rewriter} produces the final source code which is saved as a new file.

 %\systemName{} depends only on the existence of a parser for the language and a profiler that provides the performance objectives.  In current implementation, we use \acr{antlr} \cite{parr2013definitive} for generating the \acr{AST} of the program, which already supports most of the popular programming languages\footnote{\url{https://github.com/antlr/grammars-v4/}}.

\subsection{Code Transformations}

\begin{lstlisting}[aboveskip=1pt,belowskip=1pt,xleftmargin=3.5ex,label=fig:code2,style=interfaces,escapechar=|,language=Java,numbers=left,caption={Code to illustrate bad practices.}]
void func1(){
LinkedList<T> v;
v = new LinkedList<>();
v.add(new T());
int value = func3(v);
}
void func2(LinkedList<T> v){
LinkedList<T> v1 = new LinkedList<>();
int value = func3(v1);
}
int func3(LinkedList<T> v){
T t = v.get(0);
return 2*t.value;
} 
\end{lstlisting}

When implementing \systemName{}, we encountered coding practices that vastly
  increase the search space. Many turn out to be known bad
  practices~\cite{designpattern1}. Consider \autoref{fig:code2}.  In lines 2 and 8, we see two \lstinline{LinkedList} variables that
the Extractor marks \acr{Darwinian} and candidates for replacement by their
equivalent \lstinline{ArrayList} implementation.  In these lines, user is
violating the precept to "program to the interface", here \lstinline{List}, but
is, instead, declaring the variable to have the concrete, data structure not
ADT, type \lstinline{LinkedList}.  This bad practice~\cite{designpattern1} adds dependencies to the code, limiting code reuse.  They are especially
problematic for \systemName{}, because they force \systemName{} to apply
multiple transformations to replace and optimise the data structure.  Further,
\lstinline{func3} takes a \lstinline{LinkedList} as a parameter, not
\lstinline{List}, despite the fact that it only calls the \lstinline{get} method
defined by \lstinline{List} on this parameter.  This instance of violating the
"program to the interface" precept triggers a compilation error if
\systemName{} na\"ively changes \lstinline{func1}'s type. \systemName{} transforms the code to reduce the \acr{optimiser}'s search space and handle these bad practices. \systemName{} supports thee transformations - parserless, supertype, and profiler.
 
The \emph{parserless} mode changes greadily each appearance of a Darwinian implementation. When optimising \lstinline{List}, it exhaustively tries every implementation of \lstinline{List} for every \lstinline{List} variable. It is parserless, since it needs only a regular expression to identify rewritings. This makes it simple, easily portable to other languages, and fast, so it is \systemName{}' default. However, it generates invalid programs and a large search space.

\systemName{}' \emph{sypertype} transformation converts the type of a  Darwinian implementation to that of their Darwinian ADT, for example \lstinline{LinkedList<T>} $\rightarrow$ \lstinline{List<T>} on lines, $2$,$7$,$8$ and $11$. For \autoref{fig:code2}, this tranformation exposes only two DDS to the \acr{optimiser} and produces only syntactically valid code. To implement this transformation, \systemName{} invokes Eclipse's re-factoring functionality via its API, then validates the result. 
\systemName{} aims to be language-agnostic without any additional dependencies on language specific tools. 
For this case, \systemName{} auto performs this transformation by adding the supertype as an equivalent parameter in the store of data structures. 
Whenever the \acr{AST} visitor traverses a variable or parameter declaration expression it may replace the \acr{darwinian} data structure with its supertype.

"All data structures are equal, but some data structures are more equal than others" \footnote{Adapted from "Animal Farm" by George Orwell}; some DDS affect a program's performance more than others, as when one stores only a few, rarely accessed items. To rank DDS, \systemName{} \emph{profiles} its input program to identify costly methods. The \acr{Extractor} uses this info to identify the subset of a program's DDS worth considering for optimisation. \systemName{}' instrumentation is particularly important for large programs.

\subsection{Search Based Parameter Tuning}
\label{subsection:optimiser}
The \acr{optimiser} searches a combination of data structures that improves the performance of the initial program while keeps the original functionality. 
Practically, we can represent all those data structures as parameters that can be tuned using Search Based Software Engineering approaches~\cite{harman2007current}. 
Because of the nature of the various conflicting performance objectives, the problem we faced here requires a multi-objective optimisation approach to search the (near) optimal solutions.

An array of integers is used to represent the tuning parameters.
%Each integer represents a solution value for one parameter. 
Each parameter refers either to a Darwinian data structure or to the initial size of that data structure. 
If the parameter refers to a data structure, its value represents the index in the list of Darwinian data structures. 
The \acr{Optimiser} keeps additional mapping information to distinguish the types of the parameters. 
For each generation, the NSGA-II applies tournament selection, followed by a uniform crossover and a uniform mutation operation. 
In our experiments, we designed fitness functions to capture execution time, memory consumption, and CPU usage. 
After fitness evaluation, \systemName{} applies standard non-dominated selection to form the next generation. 
\systemName{} repeats this  process until the solutions in a generation converge. 
At this point, \systemName{} returns all non-dominated solutions in the final population.

\stitle{Search Space size:} We used GA because the search space is huge. Let $D$ be the definitions of darwinian data structures in program $P$. Let $I$ be the number of implementations for a particular $d \in D$. The size of the search space is:

\begin{equation}
	\prod_{d \in D}^{} I(d) * |dom(d.c)|\text{, where }d.c\text{ is }d\text{'s constructor.} 
\end{equation}

%% FIXME: the following paragraph is better in the Corpus section.
%To gather automatically the correct objectives for each subject program it is necessary that the program has some pre-defined structure. For this reason, we consider subject programs that use some built tool to provide a good basic configuration. We used Google BigQuery to analyse the public data of Github and found that the most popular build tools for Java are Maven with $111403$ usages and Grandle with $72383$ (for date 23 Feb. 2017). Currently, \systemName{} supports automatically subject programs built in Maven and we plan to integrate Grandle in the future.

% To measure the memory consumption and CPU usage of a subject program, we use the popular JConsole profiler\footnote[6]{\url{http://openjdk.java.net/tools/svc/jconsole/}} because it uses the stats directly from \acr{JDK}, and it provides an easy programmable \acr{API}. We extended JConsole to monitor only those processes that refer each test of the provided test suite. To measure the execution time of the test suite, we use the Maven Surefire plugin\footnote[7]{\url{http://maven.apache.org/components/surefire/maven-surefire-plugin/}}, which reports only the execution time of each individual test excluding the measurement overhead that other Maven plugins may introduce.

\subsection{Deployability} 
\label{subsec:deployability}
\systemName{} provides optimisation as a cloud service. To use the service, developers only need to provide the source code of their project in a Maven build format and a performance test suite invoked by \lstinline{mvn test}. \systemName{} returns the optimised source code and a performance report. \systemName{} exposes a RESTful API that developers can use to edit the default store of Darwinian data structures. The API also allows developers to select other Search Based algorithms; the \acr{Optimiser} uses NSGA-II by default. To use our tool from the command line, a simple command is used:

{\small
\captionsetup{skip=1pt}
\begin{minipage}{.9\linewidth}
\centering
\begin{lstlisting}[language=Java]
./artemis input-program-src
\end{lstlisting}
\setlength{\belowcaptionskip}{5pt}
% \captionof{lstlisting}{Data structure instrument example}\label{cf:instrument} 
\end{minipage}
}

\noindent where this command defaults to \systemName{}'s built in \acr{DDSSG}.
\systemName{} writes the source of an optimized variant of its input for each
measure.  \systemName{} also supports optional parameters to customise its
processing.

% {\footnotesize
% \begin{itemize}[\null]
% % \item \texttt{-input [foldePath]}
% % \item \texttt{-output [foldePath]}
% \item \texttt{-optimisation [data structures, parameters, both]}
% \item \texttt{-analysis [exhaustive, static, optimiser]}
% \item \texttt{-profiler [true,false]}
% \item \texttt{-limit [time, memory, CPU usage]}
% \end{itemize}
% }

% Fan says: I don't suggest listing options of the tool in the paper. This belongs to the README of the tool and at most we just need to point users to the tool's repository url.

\section{Evaluation}
\label{sec:evaluation}

To demonstrate the performance improvements that \systemName{} automatically
achieves and its broad applicability, we applied it to three corpora:  $8$
popular GitHub projects, $5$ projects from the Dacapo Benchmark, and $30$
projects, filtered to meet \systemName{}'s requirements, then sampled uniformly
at random from Github.  To show also that \systemName is language-agnostic, we
applied it to optimise
Guetzli\footnote{\url{https://github.com/google/guetzli}}~(\autoref{research_questions}), a JPEG encoder
written in C++.

\subsection{Corpus}
\label{subsec:corpus}

\systemName{} requires projects with capable build systems and an extensive
test suites.  These two requirements entail that \systemName{} be able to build
and run the project against its test suite.  \systemName{} is language-agnostic
but is currently only instantiated for Java and C++, so it requires Java or C++
programs.

Our first corpus comprises eight popular GitHub projects.  We selected these
eight to have good test suites and be diverse. 
We defined popular to be
projects that received at least $200$ stars on GitHub.  We deemed a test 
suite to be good if its line coverage met or exceeded $70$\%. 
This corpus contains projects, usually well-written, optimised and peer code-reviewed by experienced developers.
We applied \systemName{} on those projects to investigate whether it can provide a better combination of data structures than those selected by experienced human developers.

This first corpus might not be representative, precisely because of the
popularity of its benchmarks.  To address this threat to validity, we
turned to the DaCapo benchmarks~\cite{blackburn2006dacapo}.  The authors of DaCapo built 
it, from the ground up, to be representative.  The goal was to provide the research community with  realistic, large scale Java benchmarks that contain a good methodology for Java evaluation.   Dacapo contains $14$  open source, client-side Java benchmarks  (version $9.12$) and they come with built-in extensive evaluation. Each benchmark provides accurate measurements for execution time and memory consumption.
DaCapo first
appeared in 2006 to work with Java v.1.5 and has not been further updated to work with newer versions of Java.  For this reason, 
we faced difficulties in compiling all the benchmarks
and the total number of benchmarks were reduced to $5$ out of $14$. In this corpus we use the following five: \op{fop}, \op{avrora}, \op{xalan}, \op{pmd} and \op{sunflow} (\autoref{fig:dacapo}).

Because of its age and the fact that we are only using subset of it, our DaCapo
benchmark may not be representative.  To counter this threat, we uniformly sampled
projects from GitHub.  We discarded those that did
not meet \systemName{}'s constraints, like being equipped with a build system,
until we collected $30$ projects.  Those projects are diverse, both in domain
and size.  The selected projects include static analysers, testing frameworks,
web clients, and graph processing applications. Their sizes vary from $576$ to
$94K$ lines of code with a median of $14881$.  Their popularity varies from $0$
to $5642$ stars with a median of $52$ stars per project. The median number of
tests is $170$ and median line coverage ratio is $72\%$.

Collectively, we systematically built these corpora  to be representative in
order to demontrate the general applicably of the \systemName{}' optimization
framework. The full list of the programs used in this experimental study are
available online\footnote{\url{https://darwinianoptimiser.com/corpus}} in the project's website.

%We have worked hard to ensure that our overall corpus, consisting of these three subcorpora, is representative to demontrate the general applicably  of the \systemName{}' optimization framework.

% It should be clear that the popular Github projects have much more stars than the sampled Github projects (more stars means more popular). 
%Also the subjects should differ in sizfe and also in the number of available data structures that they use in order to show the benefits of a search based optimisation approach and its scalability.

\subsection{Experimental Setup}
\label{subsec:experimental_setup}
% Microsoft Azure\textsuperscript{TM} D4-v2 // it is better not to refer Azure machine because it is a virtual machine and we cannot guarantee that Microsoft is not sharing the same hardware

Experiments were conducted using Microsoft Azure\textsuperscript{TM} D4-v2 machines with one Intel E5-2673v3 CPU featuring 8 cores and 14GB of DRAM and built with Oracle JDK 1.8.0 and Ubuntu 16.04.4 LTS. 

Performance measurements may lead to incorrect results if not handled carefully~\cite{arnold2002online}. Thus, a statistical rigorous performance evaluation is required~\cite{georges2007statistically, kalibera2013rigorous,lingbo2}. To mitigate instability and incorrect results, we differentiate VM start-up and steady-state. We ran our experiments in a fresh Azure VM that contained only the JVM and the subject. 
We use JUnit, which runs an entire test suite in a single JVM. 
We manually identified and dropped startup runs, then we spot-checked the results to confirm that the rest of the runs achieved a steady state and were exhibiting low variance. 
All of the means and medians we reported fall within the computed interval with $95\%$ confidence. 
To assure the accuracy and reduce the bias in the measurement, program profiling period was set as $0.1$ seconds, and each generated solution was run for more than $30$ simulations. 
Also we use Mann Whitney U test~\cite{fay2010wilcoxon} to examine if the improvement is statistically significant.

% Ubuntu 14.04.4 LTS and Oracle JDK
% 1.8.0 have been used for development and the experiments
% have been run in a Microsoft Azure D4S-v2 virtual machine
% with one Intel E5-2673 CPU featuring 8 cores and 28 GiB
% of RAM, at a cost of £ 0.8 per hour.

To measure the memory consumption and CPU usage of a subject program, we use the popular JConsole profiler\footnote{\url{http://openjdk.java.net/tools/svc/jconsole/}} because it directly handles \acr{JDK} statistics and provides elegant \acr{API}. 
We extended JConsole to monitor only those system processes belonging to the test suite. We use 
Maven Surefire plugin\footnote{\url{http://maven.apache.org/components/surefire/maven-surefire-plugin/}} to measure the test suite's execution time because it reports only the execution time of each individual test, excluding the measurement overhead that other Maven plugins may introduce.
% To measure the memory consumption and CPU usage of a subject program, we use the popular JConsole profiler\footnote{\url{http://openjdk.java.net/tools/svc/jconsole/}} because it uses the stats directly from \acr{JDK}.  To measure the execution time of the test suite, we use the Maven Surefire plugin\footnote{\url{http://maven.apache.org/components/surefire/maven-surefire-plugin/}}.

For the \acr{Optimiser}, we chose an initial population size of $30$ and a maximum number of $900$ function evaluations. 
We used the tournament selection (based on ranking and crowding distance), simulated binary crossover (with crossover probability $0.8$) and polynomial mutation (with the mutation probability $0.1$). 
We determined these settings from  calibration trials to ensure the maturity of the results.
Since NSGA-II is stochastic, we ran each experiment $30$ times to obtain statistical significant results.
%For each program, we compared the generated Pareto-front solutions~\cite{pareto} with the initial program with respect to three performance objectives: execution time, memory consumption and CPU usage.

% \stitle{Experiment.}~To answer this question, we uniformly select $30$ sample programs from all available Github Java programs and manually $8$ famous programs that have the specific characteristics mentioned in \autoref{subsec:subjects}. We used Google BigQuery to get a list of all Java programs. Each program was run for $30$ times using $10$ simulations to eliminate any bias in the results. For each program, we compared the generated Pareto-front solutions~\cite{pareto} with the initial program with respect to three performance objectives: execution time, memory consumption and CPU usage. 

\subsection{Research Questions and Results Analysis}
\label{research_questions}
\systemName{} aims to improve all objectives at the same time. Therefore the first research question we would like to answer is:
% The question begs 'Whether \systemName{} can improve all three objectives or at least partially?'. 
% This motivates our first research question:
% In case it cannot improve all three objectives, for which type of application this does happen. 

% \begin{table}[htp]
% % \centering
% \begin{center}
% \begin{tabular}{ || c c ||  }
%  \hline
%  Characteristic & Value\\
%  \hline\hline
%  processor   & Intel 8 cores E5-2673 v3 CPU \\
% %   clock speed   & 2.5 GHz per core \\
%   memory & 14 GB 1600 MHz DDR3 \\
%   disk & 250 GB SSD \\
%   operating system & Ubuntu 16.04.4 LTS \\
%  \hline
% \end{tabular}
% \caption{Hardware characteristics.}
% \label{table:1}
% \end{center}
% \end{table}

% Add research questions
\vspace*{2mm}
\textbf{RQ1:} \textit{What proportion of programs does \systemName{} improve?}
\vspace*{2mm}

To answer RQ1, we applied \systemName{} to our corpus. We inspected the generated optimal solutions from $30$ runs of each subject by examining the dominate relation between the optimal and inital solutions regarding the optimisation objectives.
We introduce the terms \emph{strictly dominate relation} and \emph{non-dominated relation} to describe the relation.
Defined by Zitzler et al.~\cite{1197687}, a solution \emph{strictly dominates} another solution if it outperforms the latter in all measures.
A solution is \emph{non-dominated} with another solution if both outperform the other in at least one of the measures. 
% For brevity, in \autoref{table:exp1} we use \emph{strong} improvement to represent a \emph{strictly dominate relation}, while \emph{weak} improvement for a \emph{incomparable relation}.

For DaCapo, \systemName{} found at least one strictly dominant solution for $4$ out of $5$ projects; it found no such solution for \op{sunflow}. 
It found $1072$ solutions, from which $3\%$ are strictly dominant (median is $5.5$ solutions per project) and $64\%$ are non-dominated (median is $18$ solutions per project).
% weak: 690- strong: 29 bad: 4

For the popular Github projects, \systemName{} found at least one strictly dominant solution for all $8$ projects. 
The total number of solutions found is $10218$ and $16\%$ of them are strictly dominant (median is $50$ solutions per project) and $59\%$ are non-dominated (median is $749.5$ solutions per project).

For the sampled Github projects, \systemName{} found a strictly dominant solution for $25$ out of $30$ projects, but found no solution for projects \op{rubix-verifier}, \op{epubcheck}, \op{d-worker}, \op{telegrambots} and \op{fqueue}. It found $27503$ of which $10\%$ of them are strictly dominant (median is $24$ solutions per project) and $66\%$ are non dominant (median is $125$ solutions per project). With these results, we answer $RQ1$ affirmatively:
% According to the aforementioned results, we answer $RQ1$ by clearly stating that, for $37$ out of $42$ 
% \systemName{} provides strictly dominant solutions.

% weak: 24215- strong: 4420 bad: 34
% total: 28669
% percentage weak: 0.844640552513
% percentage strong: 0.154173497506
% percentage bad: 0.00118594998082
% 36
\vspace*{2mm}
\begin{mdframed}
\textbf{Finding1:} 
 \systemName{} finds optimised variants that outperform the original program in at least one measure for \emph{all} programs in our representative corpus. 
\end{mdframed}
\label{rq1}
\vspace*{2mm}
This finding understates \systemName{}'s impact.  Not only did it improve at least
one measure for \emph{all programs}, \systemName{} found solutions that improve \emph{all measures} for $88\%$ of the programs.

Having found that \systemName{} finds performance improvements, we ask "How good are these improvements" with: 

\vspace*{2mm}
\textbf{RQ2:} \textit{What is the average improvement that \systemName{} provides for each program?}
\vspace*{2mm}

Though \systemName{} aims to improve all candidate's measures, it cannot achieve that if improvements are antagonistic. In some domains, it is more important to significantly improve one of the measures than to improve slightly all measures; \eg a high frequency trading application may want to pay the cost of additional memory overhead in order to improve the execution time. 
Our intuition is that the \acr{Optimiser} will find many solutions on the Pareto-front and at least one of them will improve each measure significantly.

We answer $RQ2$ quantitatively. 
We report the maximum improvement (median value with $95\%$ confidence interval) for execution time, memory and CPU usage  for each subject of the three corpora.
We use bar charts with error bars to plot the three measures for each program. In Y axis, we represent the percentage of improvement for each measure. A value less than $100\%$ represents an improvement and a value greater than $100\%$ means degradation; \eg $70\%$ memory consumption implies that the solution consumes $70\%$ of the memory used in the input program.

\begin{figure}[tb]
	\captionsetup{width=0.45\textwidth}
	\centering
	\subfloat[Best execution time of popular GitHub programs. The median value is $93.3\%$, mean is $86.4\%$. Median number of \acr{DDS} is $12$ and mean is $14.6$. Median number of \acr{DDS} changes is $4$ and mean is $5$. \label{fig:popular1}]
	{\includegraphics[width=0.80\linewidth]{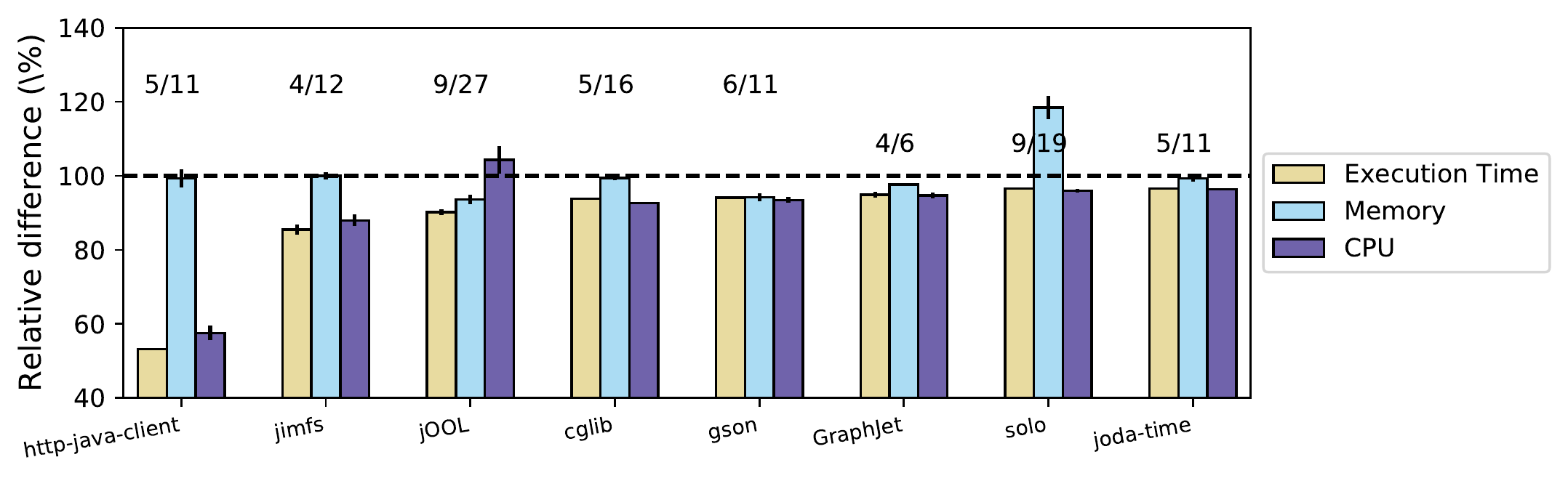}}
	
	%\\[-0.5ex]
	
	\subfloat[Best memory consumption of popular GitHub programs. The median value is $86\%$ and mean is $84\%$. Median number of \acr{DDS} is $12$ and mean is $14.6$. Median number of \acr{DDS} changes is $4$ and mean is $5.85$.\label{fig:popular2}]
	{\includegraphics[width=0.80\linewidth]{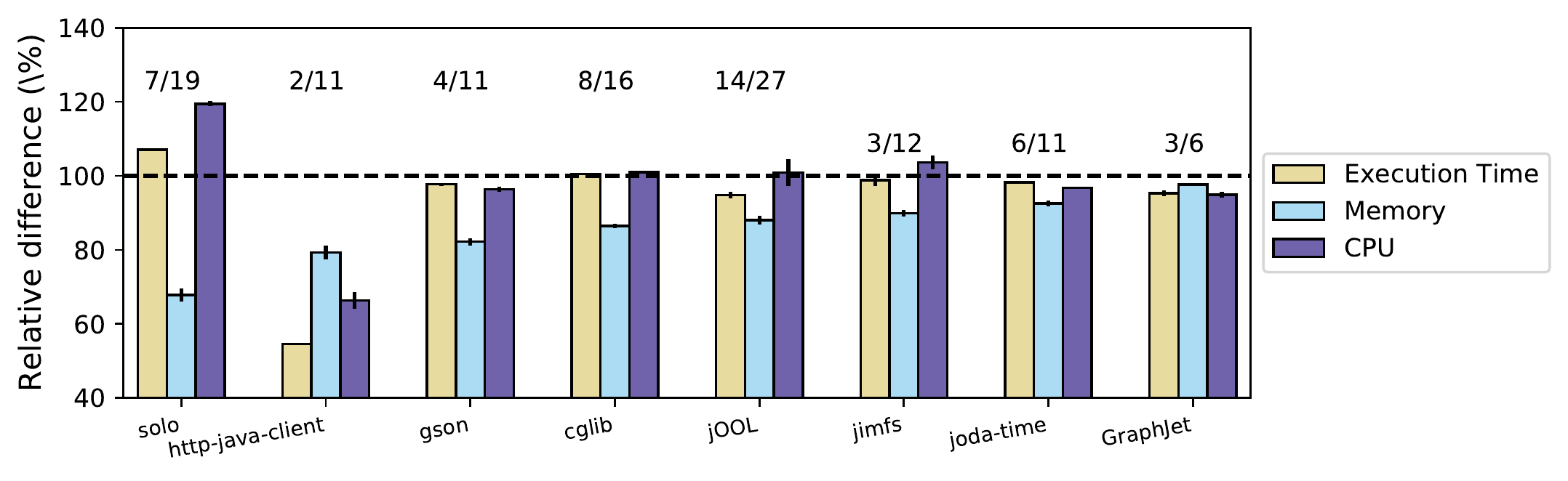}}
	
	%\\[-0.5ex]
	
	\subfloat[Best CPU usage of popular GitHub programs. The median value is $90.3\%$ and mean is $84.6\%$. Median number of \acr{DDS} is $12$ and mean is $14.6$. Median number of \acr{DDS} changes is $5$ and mean is $7.42$.\label{fig:popular3}]
	{\includegraphics[width=0.80\linewidth]{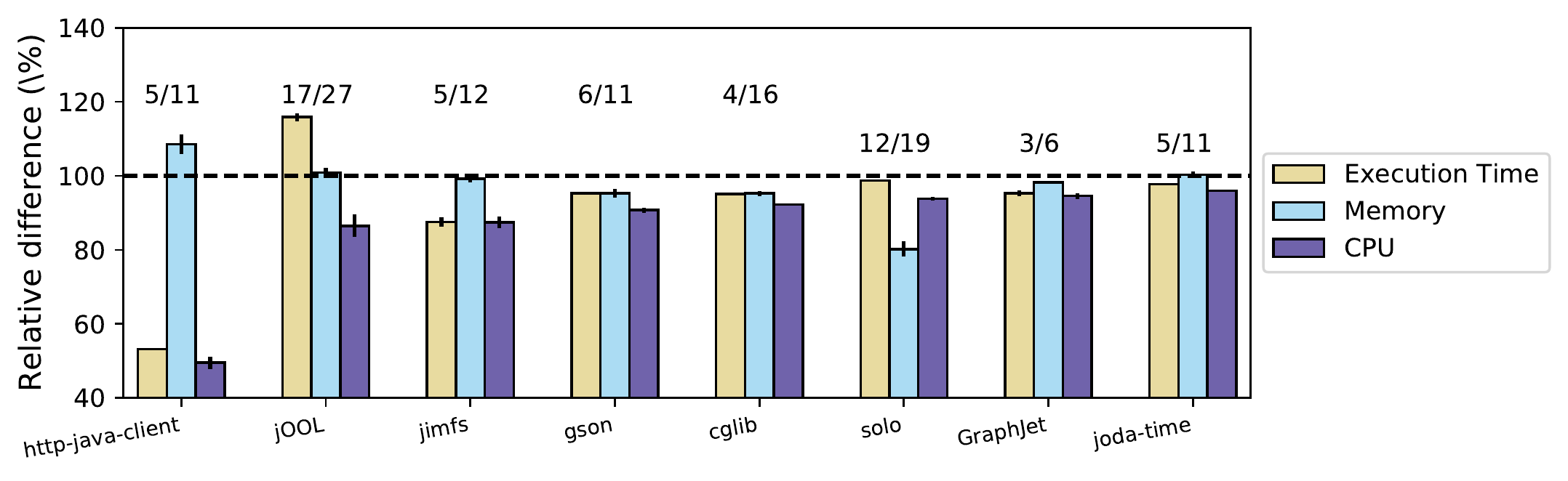}}
	
	\caption{\textbf{Answers} \textbf{\textit{RQ2}}. Description.}
	\label{fig:popular}
	\vspace{-2em}
\end{figure}

\textbf{Selected popular GitHub programs.}~\autoref{fig:popular1} presents the three measures of the solutions when the execution time is minimised, for each program from the popular GitHub programs. 
We observe that \systemName{} improves the execution time of \textit{every} program. 
 \op{google-http-java-client}'s  execution time was improved the most; its execution time was reduced by M=$46$\%, $95$\% CI [$45.6$\%, $46.3$\%]. 
We also notice that this execution time improvement did not affect negatively the other measures, but instead the CPU usage was reduced by 
M=$41.6$\%, $95$\% CI [$39.6$\%, $43.6$\%] and memory consumption remained almost the same. 
The other interesting program to notice from this graph is \op{solo}, a blogging system written in Java; its execution time improved slightly by $2$\% but its memory consumption increased by $20.2$\%. 
Finally, for this set of solutions, the median execution time improvement is $14.13$\%, whilst memory consumption slightly increased by $1.99$\% and CPU usage decreased by $3.79$\%. For those programs, \systemName{} extracted a median of $12$ data structures and the optimal solutions had a median of $4$ data structures changes from the original versions of the program.

\autoref{fig:popular2} shows the solutions optimised for memory consumption. 
We notice that \systemName{} improves the memory consumption for all programs, with a median value of $14$\%. 
The execution time was improved by a median value of $2.8$\% for these solutions, while the median value of CPU usage is slightly increased by $0.4$\%. 
We notice that \op{solo} has the best improvement by M=$31.1$\%, $95$\% CI [$29.3$\%, $33$\%], but with an increase of M=$8.7$\%, $95$\% CI [$8.5$\%, $8.9$\%] in execution time and M=$21.3$\%, $95$\% CI [$20.6$\%, $22$\%] in CPU usage. \op{Graphjet}, a real-time graph processing library, has the minimum improvement of M=$0.9$\%, $95$\% CI [$0.6$\%, $1.1$\%]. The optimal solutions had a median of $4$ data structures changes per solution.

\autoref{fig:popular3} presents solutions optimised for CPU usage. 
The median CPU usage improvement is $9.7$\%. 
The median value of execution time improved by $5.2$\% and the median value of memory consumption improved by $2.3$\%. 
The program with the most significant improvement in CPU is \op{http-java-client} with M=$49.7$\%, $95$\% CI [$48$\%, $51.4$\%], but with a decrease in memory of M=$9.8$\%, $95$\% CI [$7.5$\%, $12.9$\%]. The optimal solutions make a median of $5$ data structures changes to the original versions of the program.

\begin{figure}[tb]
\captionsetup{width=0.45\textwidth}
  \centering
  \subfloat[Best execution time of the Dacapo benchmark. The median value is $95.20\%$ and mean is $95.6\%$. Median number of \acr{DDS} is $18$ and mean is $14.8$. Median number of \acr{DDS} changes is $5$ and mean is $4.8$.\label{fig:dacapo1}]
  {\includegraphics[width=0.70\linewidth]{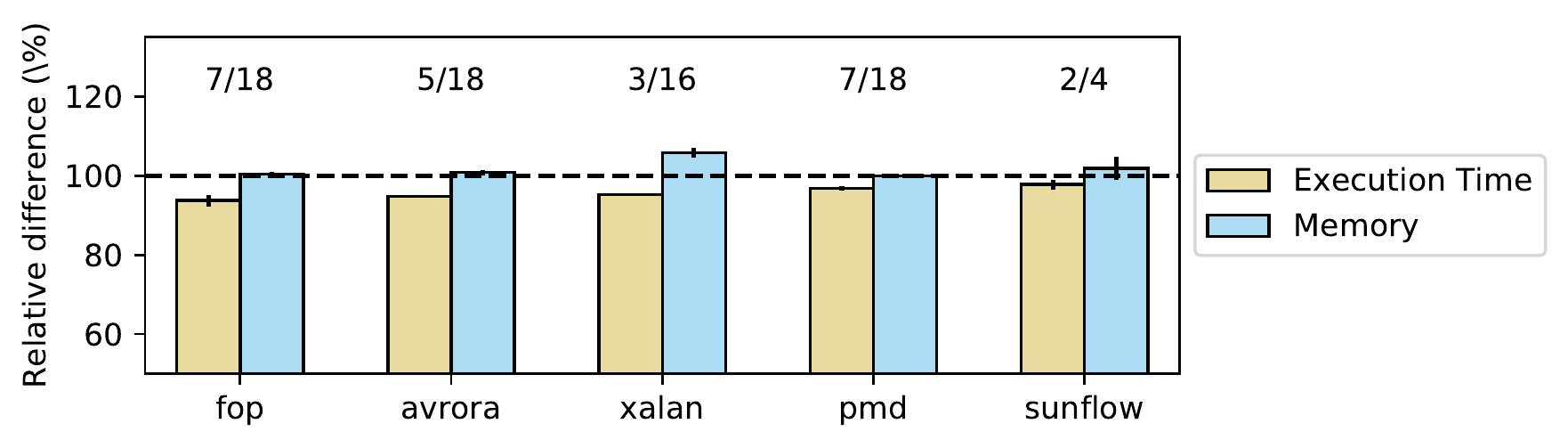}}
  
  %\\[-0.5ex]
  
  \subfloat[Best memory consumption of the Dacapo benchmark. The median value is $95.7\%$ and mean is $92.1\%$ . Median number of \acr{DDS} is $18$ and mean is $14.8$. Median number of \acr{DDS} changes is $3$ and mean is $4.2$.\label{fig:dacapo2}]
  {\includegraphics[width=0.70\linewidth]{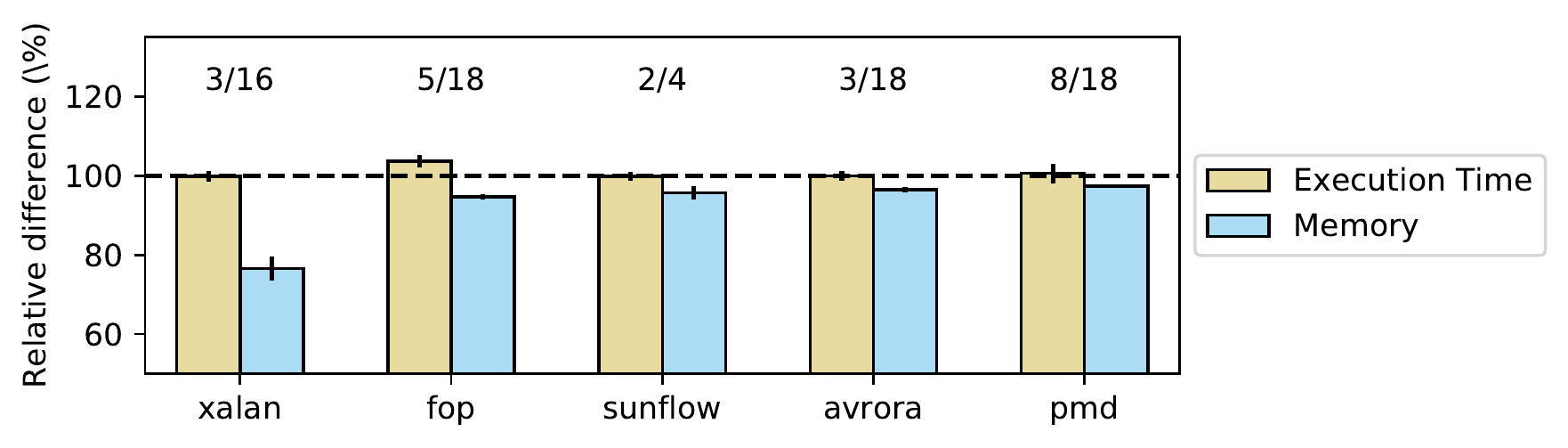}}

  \caption{\textbf{Answers} \textbf{\textit{RQ2}}. Description.}
  \label{fig:dacapo}
  \vspace{-3.5em}
\end{figure}

\textbf{DaCapo.} \autoref{fig:dacapo} presents all solutions optimised for execution time and memory consumption for the DaCapo benchmark. We used only two measures for the DaCapo benchmark as those were the ones built in the benchmark suite. We chose not to extend or edit the profiling method of DaCapo, to avoid the risk of affecting the validity of its existing, well tested profiling process.

\systemName{} found solutions that improve the execution time for every program without affecting significantly the memory consumption, except project \op{xalan} which had improvement  (M=$4.8$\%, 95\% CI [$4.6$\%, $5.7$\%] in execution time but with an increase ($5.8$\%, 95\% CI [$3.5$\%, $7$\%]) in memory consumption. All solutions for optimised memory consumption did not affect execution time, except for a slight increase for program \op{fop}. Finally, for this set of solutions, the median percentage of execution time improvement is $4.8$\%, and $4.6$\% for memory consumption.  For this set of programs, \systemName{} extracted a median of $18$ data structures per program, and the optimal solutions had a median of $5$ data structures changes for the execution time optimised solutions and $4$ for the memory optimised solutions.

\begin{figure*}

\begin{minipage}{1\linewidth}
\centering
\captionsetup{width=0.80\textwidth}
\subfloat[Best execution time of uniformly selected GitHub programs. The median value is $95.4\%$ and mean is $94.7\%$. Median number of \acr{DDS} is $9.5$ and mean is $11.6$. Median number of \acr{DDS} changes is $5$ and mean is $4.8$.]{\includegraphics[scale=0.20]{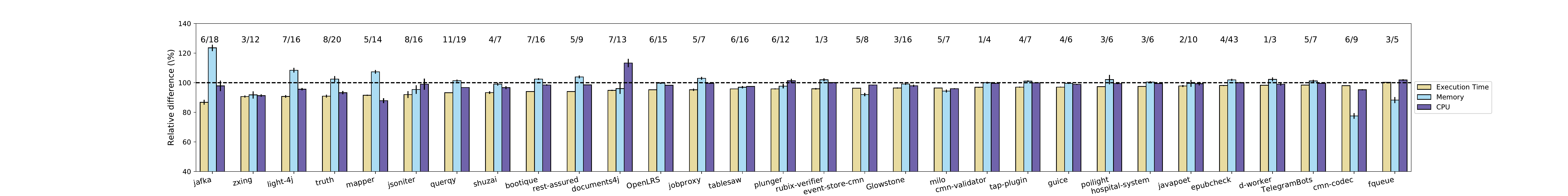}\label{fig:uniform1}}
\end{minipage}

\begin{minipage}{1\linewidth}
\centering
\captionsetup{width=0.80\textwidth}
\subfloat[Best memory consumption of the uniformly selected GitHub programs. The median value is $89.1\%$ and mean is $86.8\%$. Median number of \acr{DDS} is $9.5$ and mean is $11.6$. Median number of \acr{DDS} changes is $5$ and mean is $4.6$.]{\includegraphics[scale=0.20]{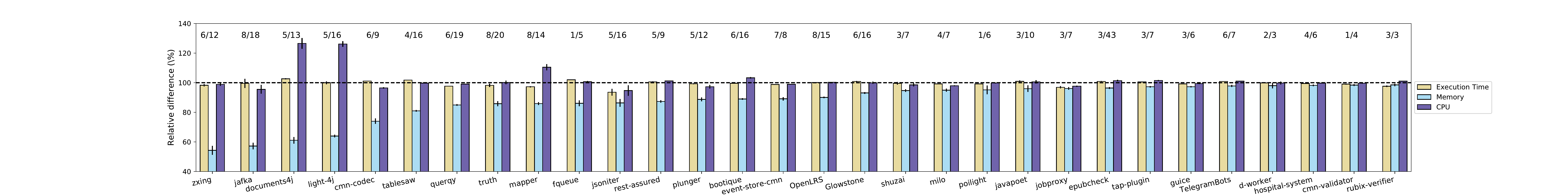}\label{fig:uniform2}}
\end{minipage}

\begin{minipage}{1\linewidth}
\centering
\captionsetup{width=0.80\textwidth}
\subfloat[Best CPU usage of the uniformly selected GitHub programs. The median value is $5.1\%$ and mean is $8\%$. Median number of \acr{DDS} is $9.5$ and mean is $11.6$. Median number of \acr{DDS} changes is $5$ and mean is $4.5$.]{\includegraphics[scale=0.20]{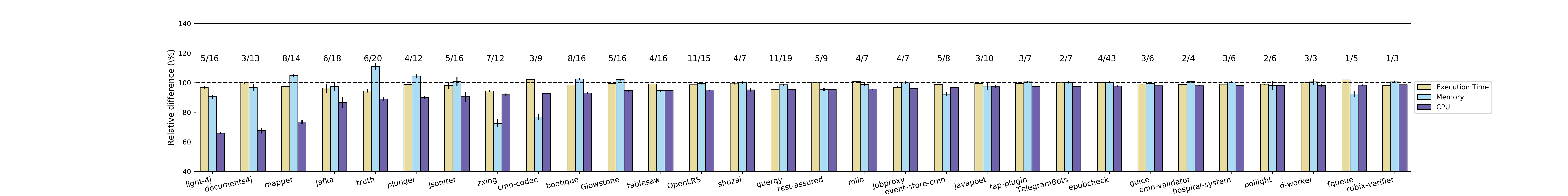}\label{fig:uniform3}}
\end{minipage}

\caption{\textbf{Answers} \textbf{\textit{RQ2}}. Description.}
\label{fig:uniform}
\vspace{-1.5em}
\end{figure*}

% \begin{figure*}
% \begin{minipage}{1\linewidth}
% \centering
% \subfloat[Best execution time of uniformly selected GitHub programs. The median value is $92.04\%$.]{\includegraphics[scale=0.52]{figures/RQ2_1_1.pdf}\label{fig:uniform1}}
% \end{minipage}

% \begin{minipage}{1\linewidth}
% \centering
% \subfloat[Best memory consumption of the uniformly selected GitHub programs. The median value is $82.63\%$.]{\includegraphics[scale=0.52]{figures/RQ2_2_1.pdf}\label{fig:uniform2}}
% \end{minipage}

% \begin{minipage}{1\linewidth}
% \centering
% \subfloat[Best CPU usage of the uniformly selected GitHub programs. The median value is $92.22\%$.]{\includegraphics[scale=0.52]{figures/RQ2_3_1.pdf}\label{fig:uniform3}}
% \end{minipage}

% \caption{\textbf{Answers} \textbf{\textit{RQ2}}. Description.}
% \label{fig:uniform}
% \end{figure*}

\stitle{Sampled GitHub programs.}~\autoref{fig:uniform} presents all solutions optimised for execution time, memory consumption and CPU usage for the sampled GitHub programs. As with the previous corpora, \systemName{} found solutions that improved each measure significantly. \systemName{} improves the median value of execution time across all projects by $4.6$\%, memory consumption by $11.4$\% and CPU usage by $4.6$\%.

\systemName{} found solutions with antagonistic improvement for projects \op{jafka} and \op{documents4j}. \systemName{} found a solution that improves the execution time of \op{jafka}, a distributed publish-subscribe messaging system, by M=$12$\%, $95$\% CI [$11.2$\%, $13.6$\%], but also increases its memory consumption by M=$23.6$\%, $95$\% CI [$21.4$\%, $25.7$\%]. It also found a solution that improves the memory consumption of \op{documents4j} (M=$38$\%, $95$\% CI [$38$\%, $41$\%]) but introduced extra CPU usage M=$26.1$\%, $95$\% CI [$24.2$\%, $28$\%]. A median of $9.5$ data structures were extracted and the optimal solutions had a median of $5$ data structures changes from the original versions of the program.

Observing again the numbers across the three corpora, we can say that they are quite consistent, showing that \systemName{} finds optimal solutions that improve significantly the different optimisation measures. We also see that the number of Darwinian Data structures extracted (between $9.5$ and $18$)  and the optimal solutions DDS changes (between $4$ and $5$) are quite similar for the three corpora.

Analysing all results from the $3$ corpora we conclude the discussion of RQ2 with:

\vspace*{2mm}
\begin{mdframed}
\textbf{Finding2:}  \systemName{} improves the median across all programs in our  corpus by $4.8\%$ execution time, $10.2\%$ memory consumption, and $5.1\%$ CPU usage. 
\end{mdframed}
\vspace*{2mm}
% \op{light-4j\char`-http\char`-java\char`-client}

% 66.9167694614,66.3399998222,67.4935391007

\vspace*{2mm}
\textbf{RQ3:} \textit{Which Darwinian data structures does \systemName{} find and tune?}
\vspace*{2mm}

We ask this question to understand which changes  \systemName{} makes to a program. 
 \autoref{tb:subjects} contains the transformations \systemName{} applied  across all optimal solutions.  We see that the most common transformation for all measures is replacing \lstinline{ArrayList} with \lstinline{LinkedList}, it appears $91$, $86$ and $87$ times respectevely across all measures. This transformation indicates that most developers prefer to use \lstinline{ArrayList} in their code, which in general is considered to be faster, neglecting use cases in which \lstinline{LinkedList} performs better; \eg{} when the program has many list insertion or removal operations. 
 Except \lstinline{HashMap} to \lstinline{LinkedHashMap}, the other transformations happen relatively rare in the optimal solutions.  Last, the median number of lines Artemis changes is $5$.

% The main diffr LinkedList implements it with a doubly-linked list. ArrayList implements it with a dynamically re-sizing array.
% For instance, \systemName{} replaced HashMap with LinkedHashMap in $42$ of the $135$ changes it made across all optimal solutions.

% To minimise the search space, in projects that the number of DDS was bigger than $16$ we applied \systemName{} only to the most used code in as identified by the preprocessor. 

\begin{table}[tb]
\captionsetup{font=small}
\centering
\resizebox{0.49\textwidth}{!}{
\begin{tabular}{@{}lrrrr@{}}
\toprule
 Tranformation              & Time & Memory & CPU \\ \midrule
HashMap             -> LinkedHashMap        & 60   & 53     & 57  \\
LinkedList          -> ArrayList            & 16   & 13     & 18  \\
HashSet             -> LinkedHashSet        & 22   & 21     & 21  \\
LinkedBlockingQueue -> LinkedTransferQueue  & 1    & 2      & 2   \\
ArrayList           -> LinkedList           & 91   & 86     & 87  \\
LinkedHashSet       -> HashSet              & 7    & 8      & 5   \\
Vector              -> CopyOnWriteArrayList & 1    & 0      & 2   \\
LinkedHashMap       -> HashMap              & 17   & 23     & 19  \\ \bottomrule
\end{tabular}
}
\caption{DDS changes for optimal solutions across all measures.}
\label{tb:subjects}
\vspace{-4em}
\end{table}

\vspace*{2mm}
\begin{mdframed}
\textbf{Finding3:} \systemName{} extracted a median of $12$ Darwinian data structures from each program and the optimal solutions had a median of $5$ data structure changes from the original versions of the program.
\end{mdframed}
\vspace*{2mm}

\vspace*{2mm}
\textbf{RQ4:} \textit{What is the cost of using \systemName{}}?
\vspace*{2mm}

% \textit{ What is the cost of using \systemName{}?}

In order for \systemName{} to be practical and useful in real-world situations, it is important to understand the cost of using it. 
The aforementioned experimental studies reveal that, even for the popular programs, the existing selection of the data structure and the setting of its parameters may be sub-optimal. 
Therefore, optimising the data structures and their parameters can still provide significant improvement on non-functional properties. To answer this research question, the cost of \systemName{} for optimising a program is measured by the cost of computational resources it uses.
In this study, we used a Microsoft Azure\textsuperscript{TM} D4-v2 machine, which costs \textsterling $0.41$ per hour at a standard Pay-As-You-Go rate\footnote{\url{https://azure.microsoft.com/en-gb/pricing/}}, to conduct all experiments.
%Estimating developer's time to perform this optimisation is a tricky task because it depends on many factors. 
%However, we will try to do a simple estimation based on the number of data structures that need to be optimised and an empirical estimation of the time that someone needs to spend for optimising a data structure. 
%The average number of darwinian data structures that \systemName{} found for all projects was $18$; note that the real number of all data structures is probably much bigger. Let's assume that a developer finds with the first attempt all those data structure and needs around $2$ hours to try different combinations to decide which is the optimised version. Thus, a developer needs to spend around $36$ hours (or $5$ working days) in the best case scenario. In our experiments, \systemName{} did not need to run more than two days for any of the projects.

% Assume we need to rent one machine from Microsoft Cloud. 
%The price for renting an Azure\footnote[9]{ \url{https://github.com/apache/commons-collections}} virtual machine with $4$ cores, $8$GB of Ram and $40$GB of SSD disk is $3.41$ GBP per day much cheaper than the minimum price an organisation has to pay for a software engineer; the average software engineer salary for year 2017 in UK is $70$£ per day. Thus, it is obvious that the cost of \systemName{} for this optimisation process is much cheaper than any manual process.

The experiments show that an optimisation process takes $3.05$ hours on average for all studied subjects.
The program \op{GraphJet} and \op{jimfs} are the most and the least time-consuming programs respectively, with $19.16$ hours and $3.12$ minutes optimisation time.
% \systemName{} spent $19.16$ hours on average to optimise the program \op{GraphJet} whilst only $3.12$ minutes for program \op{jimfs}.
Accordingly, the average cost of applying \systemName{} for the subjects studied is \textsterling $1.25$, with a range from \textsterling $0.02$ to \textsterling $7.86$.
The experimental results show that overall cost of using \systemName{} is negligible compared to a human software engineer, with the assumption that a competent software engineer can find those optimisation in a reasonable time.

 \systemName{} transforms the selection of data structure and sets parameters by rewriting source code, 
thereby allowing human developers to easily investigate its changes and gain insight about the usage of data structures and the characteristics of the program.

%The real cost benefit from using \systemName{} comes when the program is used for products that the latency really matters. There have been a lot of stories~\cite{rq3story1,rq3story2, rq3story3} about small latencies (milliseconds) in getting search results or processing data that impacted significantly the profits of an organisation. In our corpus, we optimised similar projects that may affect significantly the core business of an organisation; \eg optimising the \op{Cleo} project which is used in the auto-complete search feature may affect the profits of LinkedIn as search is a very important feature of their platform.  

\vspace*{2mm}
\begin{mdframed}
\textbf{Finding4:} The cost of using \systemName{} is negligible, with an average of \textsterling $1.25$ per project, providing engineers with insights about the optimal variants of the project under optimisation.
\end{mdframed}
\vspace*{2mm}
%and the gains it may offer can be huge and long term.

% \cite{rq3story1, rq3story2}

% Cost that our tool can save if used for a tool that is used in products where latency matters\mb{this sentence needs to be re-written}. We can argue about Jafka in our experiments.

% Finally, the cost benefit of using \systemName{} \mb{something seems to be missing}

% https://www.fastcompany.com/1825005/how-one-second-could-cost-amazon-16-billion-sales
% http://blog.gigaspaces.com/amazon-found-every-100ms-of-latency-cost-them-1-in-sales/

% @misc{story1,
%   author = {Patrick Wyatt},
%   title = {{Avoiding game crashes related to linked lists}},
%   howpublished = "\url{http://www.codeofhonor.com/blog/avoiding-game-crashes-related-to-linked-lists}",
%   year = {2012}, 
%   note = "[Online; accessed 18-February-2017]"
% }

To show the versatility of the \systemName{} framework, we ask RQ2, RQ3 and RQ4 over Google \op{guetzli}, a very popular JPEG encoder written in C++. We used the STL containers and their operations as Darwinian data structures. More specifically, we considered the \lstinline{push_back} and \lstinline{emplace_back} as equivalent implementations of the same functionality and exposed those as tunable parameters to \systemName{}'s optimiser. We collected a random sample of images (available online~\footnote{\url{http://darwinianoptimiser.com/corpus}}) and used it to construct a performance suite that evaluates the execution time of \op{guetzli}. 

% \systemName{} extracted and tuned $25$ parameters and found an optimal solution with $11$ parameter changes that improves execution time by $7\%$.

We answer RQ2 by showing that \systemName{} found an optimal solution that improves execution time by $7\%$. We answer RQ3 by showing that \systemName{} extracted and tuned $25$ parameters and found an optimal solution with $11$ parameter changes. \systemName{} spent $1.5$ hours (costs \textsterling $0.62$) to find optimal solutions which is between the limits reported in RQ4. Last, we spent approximately $4$ days to extend \systemName{} to support C++, using the \emph{parserless} mode.

\section{Threats to Validity}
\label{sec:threats}

\autoref{subsec:corpus} discusses the steps we took to address the threats to the external validity of the results we present here.  
In short, we built three subcorpora, each more representative than the last, for a total of $43$ programs, diverse in size and domain.  
The biggest threat to the internal validity of our work is the difficulty of taking accurate performance measurements of applications running on VM, like the JVM.  \autoref{subsec:experimental_setup} details the steps, drawn from best practice, we took to address this threat.
In essence, we conducted calibration experiments to adjust the parameters such that the algorithm converges quickly and stops after the results become stable.
For measuring the non-functional properties, we carefully chose JConsole profiler that directly gathers runtime information from \acr{JDK}, such that the measurement error is minimised.
Moreover, we carefully tuned JConsole to further improve the precision of the measurements by maximising its sampling frequency such that it does not miss any measurements while minimising the CPU overhead.
To cater for the stochastic nature of \systemName{} and to provide the statistic power for the results, we ran each experiment $30$ times and manually checked that experiments had a steady state and exhibited low variance. 

\section{Related Work}\label{sec:related_work}

Multi-objective Darwinian Data Structure selection and optimisation stands
between two areas: search-based software engineering and data structure
performance optimisation.

\subsection{Search-based software engineering} 
Previous work applies Genetic
Programming~\cite{poli2008field,lingbo6,petke2017genetic,lingbo1,fan1,fan2} to either improve the functionality (bug
fixing)~\cite{6227211,lingbo5} or non-functional properties of a
program~\cite{Bruce:2015:REC:2739480.2754752,petke2014using,langdon2014improving,fan1,fan2,fan3}.
Their approaches use existing code as the code base and replace some of the
source code in the program under optimisation with the code from the code base.
However, many of these approaches rely on the Plastic Surgery
Hypothesis~\cite{Barr:2014:PSH:2635868.2635898}, which assumes that the
solutions exist in the code base. \systemName{}, on the other hand, does not
rely on this hypothesis.  \systemName{} can harvest Darwinian data structures
both from the program, but also from external code repositiories; further,
\systemName{} relies on a set of transformation rules that it can automatically
exhaustively extract from library code or documentation.

Wu et al.~\cite{fan2015} introduced a mutation-based method to expose ``deep''
parameters, similar to those we optimise in this paper, from the program under
optimisation, and tuned these parameters along with ``shallow'' parameters to
improve the time and memory performance of the program.  Though the idea of
exposing additional tunable parameter is similar to \systemName{}, their
approach did not optimise data structure selection, which can sometimes be more
rewarding than just tuning the parameters.  Moreover, they applied their
approach to a memory management library to benefit that library's clients.
 The extent of improvement usually depends on how much a program
relies on that library.  In contrast, \systemName{} directly applies to the
source code of the program, making no assumptions about which libraries the
program uses, affording \systemName{} much wider applicability.

\subsection{Data structure optimisation and bloat}
A body of work \cite{bloat1,bloat2,bloat3,bloat4,bloat5,nagel2017self,basios2017optimising} has attempted to identify bloat arising from data structures. In 2009, Shacham et al.~\cite{Shacham:2009:CAS:1542476.1542522,Shacham:2009:CAS:1543135.1542522} introduced a semantic profiler that provides online collection-usage semantics for Java programs. 
They instrumented Java Virtual Machine (JVM) to gather the usage statistics of collection data structures. 
Using heuristics, they suggest a potentially better choice for a data structure for a program.

 Though developers can add heuristics, if they lack sufficient knowledge about the data structures, they may bias the heuristics and jeopardise the effectiveness of the approach. 
\systemName{} directly uses the performance of a data structure profiled against a set of performance tests to determine the optimal choices of data structures. 
Therefore, \systemName{} does not depend on expert human knowledge about the internal implementation and performance differences of data structures to formulate heuristics. Instead. \systemName{} relies on  carefully-chosen performance tests to minimse bias. 
Furthermore, \systemName{} directly modifies the program instead of providing hints, thus users can use the fine-tuned program \systemName{}  generates without any additional manual adjustment.

 Other frameworks  provide users with manually or automatically generated selection heuristics to improve the data structure selection process. JitDS~\cite{jitds} exploits declarative specifications embedded by experts in data structures to adapt them. CollectionSwitch~\cite{costa2018collectionswitch} uses data and user-defined performance rules to select other data structure variants. Brainy~\cite{jung2011brainy} provides users with machine learning cost models that guide the selection of data structures. \systemName{} does not require expert annotations, user-defined rules or any machine learning knowledge. Storage strategies~\cite{storagestrategies} changes VMs to optimize their performance on collections that contain a single primitive type;  Artemis rewrites source code and handles user-defined types and does not make VM modifications.  

In 2014, Manotas et al.~\cite{Manotas:2014:SSE:2568225.2568297} introduced a collection data structure replacement and optimisation framework named \emph{SEEDS}. 
Their framework replaces the collection data structures in Java applications with other data structures exhaustively and automatically select the most energy efficient one to improve the overall energy performance of the application. 
%However, their framework failed to handle the additional parameters given to those data structures. 
Conceptually \systemName{} extends this approach to optimise both the data structures and their initialization parameters.
\systemName{} also extends the optimisation objectives from single objective to triple objectives and used Pareto non-dominated solutions to show the trade-offs between these objectives.
Due to a much larger search space in our problem, the exhaustive exploration search that used by \emph{SEEDS} is not practical, therefore we adopted meta-heuristic search.

Furthermore, \systemName{} directly transforms the source code of the programs
whilst \emph{SEEDS} transforms the bytecode, so \systemName{} provides developers more intuitive information about what was changed and
teaches them to use more efficient data structures. Moreover,
\systemName{} can be more easily applied to other languages as it does not
depend on language specific static analysers and refactoring tools such as
WALA~\cite{wala} and Eclipse IDE's refactoring tools. In order to support
another language we just need the grammar of that language
and to implement a visitor that extracts a program's Darwinian data structures.
We note that \acr{Antlr}, which \systemName{} uses, 
currently provides many available grammar
languages~\footnote{\url{https://github.com/antlr/grammars-v4/}}.

Apart from the novelties mentioned above, this is the largest empirical study to our knowledge compared to similar work. In the studies mentioned above, 
only $4$ to $7$ subjects were included in the experiments. Our study included the DaCapo benchmark, $30$ sampled Github subjects and $8$ well-written popular subjects to show the effectiveness of \systemName{}, therefore our results are statistically more meaningful.

\section{Conclusion}\label{sec:conclusions}

Developers frequently use underperformed data structures and forget to optimise them with respect to some critical non-functional properties once the functionalities are fulfilled. 
In this paper, we introduced \systemName{}, a novel multi-objective multi-language search-based framework that automatically selects and optimises Darwinian data structures and their arguments in a given program. 
\systemName{} is language agnostic, meaning it can be easily adapted to any programming language; extending \systemName{} to support C++ took approximately $4$ days. 
Given as input a data structure store with Darwinian implementations, it can automatically detect and optimise them along with any additional parameters to improve the non-functional properties of the given program. 
In a large empirical study on $5$ DaCapo benchmarks, $30$ randomly sampled projects and $8$ well-written popular Github projects, \systemName{} found \emph{strong} improvement for all of them.
On extreme cases, \systemName{} found $46\%$ improvement on execution time, $44.9\%$ improvement on memory consumption, and $49.7\%$ improvement on CPU usage.  \systemName{} found such improvements making small changes in the source code; the median number of lines \systemName{} changes is $5$.  Thus, \systemName{} is practical and can be easily used on other projects.
At last, we estimated the cost of optimising a program in machine hours.
With a price of \textsterling $0.41$ per machine hour, the cost of optsimising any subject in this study is less than \textsterling $8$, with an average of \textsterling $1.25$.
Therefore, we conclude that \systemName{} is a practical tool for optimising the data structures in large real-world programs.

\section*{Acknowledgements}
We would like to thank Graham Barrett, David Martinez, Kenji Takeda and Nick Page for  their invaluable assistance with respect to developing \systemName{}. Lastly, we are grateful to Microsoft Azure and Microsoft Research for the resources and commercial support.
%\emph{Omitted to meet the double-blind review requirements of FSE.}

\bibliographystyle{abbrv}
\bibliography{bibliography}

\begin{thebibliography}{10}

\bibitem{arnold2002online}
M.~Arnold, M.~Hind, and B.~G. Ryder.
\newblock Online feedback-directed optimization of java.
\newblock In {\em ACM SIGPLAN Notices}, volume~37, pages 111--129. ACM, 2002.

\bibitem{Barr:2014:PSH:2635868.2635898}
E.~T. Barr, Y.~Brun, P.~Devanbu, M.~Harman, and F.~Sarro.
\newblock The plastic surgery hypothesis.
\newblock In {\em Proceedings of the 22Nd ACM SIGSOFT International Symposium
  on Foundations of Software Engineering}, FSE 2014, pages 306--317, New York,
  NY, USA, 2014. ACM.

\bibitem{basios2017optimising}
M.~Basios, L.~Li, F.~Wu, L.~Kanthan, and E.~T. Barr.
\newblock Optimising darwinian data structures on google guava.
\newblock In {\em International Symposium on Search Based Software
  Engineering}, pages 161--167. Springer, 2017.

\bibitem{binder2000testing}
R.~V. Binder.
\newblock {\em Testing object-oriented systems: models, patterns, and tools}.
\newblock Addison-Wesley Professional, 2000.

\bibitem{blackburn2006dacapo}
S.~M. Blackburn, R.~Garner, C.~Hoffmann, A.~M. Khang, K.~S. McKinley,
  R.~Bentzur, A.~Diwan, D.~Feinberg, D.~Frampton, S.~Z. Guyer, et~al.
\newblock The dacapo benchmarks: Java benchmarking development and analysis.
\newblock In {\em ACM Sigplan Notices}, volume~41, pages 169--190. ACM, 2006.

\bibitem{storagestrategies}
C.~F. Bolz, L.~Diekmann, and L.~Tratt.
\newblock Storage strategies for collections in dynamically typed languages.
\newblock In {\em Proceedings of the 2013 ACM SIGPLAN International Conference
  on Object Oriented Programming Systems Languages \&\#38; Applications},
  OOPSLA '13, pages 167--182, New York, NY, USA, 2013. ACM.

\bibitem{brown2010managing}
N.~Brown, Y.~Cai, Y.~Guo, R.~Kazman, M.~Kim, P.~Kruchten, E.~Lim,
  A.~MacCormack, R.~Nord, I.~Ozkaya, et~al.
\newblock Managing technical debt in software-reliant systems.
\newblock In {\em Proceedings of the FSE/SDP workshop on Future of software
  engineering research}, pages 47--52. ACM, 2010.

\bibitem{Bruce:2015:REC:2739480.2754752}
B.~R. Bruce, J.~Petke, and M.~Harman.
\newblock Reducing energy consumption using genetic improvement.
\newblock In {\em Proceedings of the 2015 Annual Conference on Genetic and
  Evolutionary Computation}, GECCO '15, pages 1327--1334, New York, NY, USA,
  2015. ACM.

\bibitem{costa2018collectionswitch}
D.~Costa and A.~Andrzejak.
\newblock Collectionswitch: a framework for efficient and dynamic collection
  selection.
\newblock In {\em Proceedings of the 2018 International Symposium on Code
  Generation and Optimization}, pages 16--26. ACM, 2018.

\bibitem{dale1996abstract}
N.~Dale and H.~M. Walker.
\newblock {\em Abstract data types: specifications, implementations, and
  applications}.
\newblock Jones \& Bartlett Learning, 1996.

\bibitem{lingbo5}
H.~Dan, M.~Harman, J.~Krinke, L.~Li, A.~Marginean, and F.~Wu.
\newblock Pidgin crasher: searching for minimised crashing gui event sequences.
\newblock In {\em International Symposium on Search Based Software
  Engineering}, pages 253--258. Springer, 2014.

\bibitem{jitds}
M.~De~Wael, S.~Marr, J.~De~Koster, J.~B. Sartor, and W.~De~Meuter.
\newblock Just-in-time data structures.
\newblock In {\em 2015 ACM International Symposium on New Ideas, New Paradigms,
  and Reflections on Programming and Software (Onward!)}, pages 61--75. ACM,
  2015.

\bibitem{nsgaii}
K.~Deb, A.~Pratap, S.~Agarwal, and T.~Meyarivan.
\newblock A fast and elitist multiobjective genetic algorithm: Nsga-ii.
\newblock {\em IEEE transactions on evolutionary computation}, 6(2):182--197,
  2002.

\bibitem{bloat3}
B.~Dufour, B.~G. Ryder, and G.~Sevitsky.
\newblock A scalable technique for characterizing the usage of temporaries in
  framework-intensive java applications.
\newblock In {\em Proceedings of the 16th ACM SIGSOFT International Symposium
  on Foundations of software engineering}, pages 59--70. ACM, 2008.

\bibitem{fay2010wilcoxon}
M.~P. Fay and M.~A. Proschan.
\newblock Wilcoxon-mann-whitney or t-test? on assumptions for hypothesis tests
  and multiple interpretations of decision rules.
\newblock {\em Statistics surveys}, 4:1, 2010.

\bibitem{georges2007statistically}
A.~Georges, D.~Buytaert, and L.~Eeckhout.
\newblock Statistically rigorous java performance evaluation.
\newblock {\em ACM SIGPLAN Notices}, 42(10):57--76, 2007.

\bibitem{6227211}
C.~L. Goues, M.~Dewey-Vogt, S.~Forrest, and W.~Weimer.
\newblock A systematic study of automated program repair: Fixing 55 out of 105
  bugs for \$8 each.
\newblock In {\em 2012 34th International Conference on Software Engineering
  (ICSE)}, pages 3--13, June 2012.

\bibitem{story3}
B.~Hardin.
\newblock {Companies with hacking cultures fai}.
\newblock
  \url{https://blog.bretthard.in/companies-with-hacking-cultures-fail-b8907a69e3d#.ffdkyb1w2},
  2016.
\newblock [Online; accessed 25-February-2017].

\bibitem{harman2007current}
M.~Harman.
\newblock The current state and future of search based software engineering.
\newblock In {\em 2007 Future of Software Engineering}, pages 342--357. IEEE
  Computer Society, 2007.

\bibitem{wala}
IBM.
\newblock {T.J. Watson Libraries for Analysis (WALA).}
\newblock \url{http://wala.sourceforge.net/wiki/index.php/Main_Page}, 2009.
\newblock [Online; accessed 18-February-2017].

\bibitem{fan1}
Y.~Jia, F.~Wu, M.~Harman, and J.~Krinke.
\newblock Genetic improvement using higher order mutation.
\newblock In {\em Genetic and Evolutionary Computation Conference, {GECCO}
  2015, Madrid, Spain, July 11-15, 2015, Companion Material Proceedings}, pages
  803--804, 2015.

\bibitem{jung2011brainy}
C.~Jung, S.~Rus, B.~P. Railing, N.~Clark, and S.~Pande.
\newblock Brainy: effective selection of data structures.
\newblock In {\em ACM SIGPLAN Notices}, volume~46, pages 86--97. ACM, 2011.

\bibitem{kalibera2013rigorous}
T.~Kalibera and R.~Jones.
\newblock Rigorous benchmarking in reasonable time.
\newblock In {\em ACM SIGPLAN Notices}, volume~48, pages 63--74. ACM, 2013.

\bibitem{Knuth:1974:SPG:356635.356640}
D.~E. Knuth.
\newblock Structured programming with go to statements.
\newblock {\em ACM Comput. Surv.}, 6(4):261--301, Dec. 1974.

\bibitem{langdon2014improving}
W.~B. Langdon, M.~Modat, J.~Petke, and M.~Harman.
\newblock Improving 3d medical image registration cuda software with genetic
  programming.
\newblock In {\em Proceedings of the 2014 Annual Conference on Genetic and
  Evolutionary Computation}, pages 951--958. ACM, 2014.

\bibitem{lingbo4}
L.~Li.
\newblock Exact analysis for next release problem.
\newblock In {\em Requirements Engineering Conference (RE), 2016 IEEE 24th
  International}, pages 438--443. IEEE, 2016.

\bibitem{lingbo3}
L.~Li.
\newblock {\em Exact analysis for requirements selection and optimisation}.
\newblock PhD thesis, UCL (University College London), 2017.

\bibitem{lingbo2}
L.~Li, M.~Harman, E.~Letier, and Y.~Zhang.
\newblock Robust next release problem: handling uncertainty during
  optimization.
\newblock In {\em Proceedings of the 2014 Annual Conference on Genetic and
  Evolutionary Computation}, pages 1247--1254. ACM, 2014.

\bibitem{lingbo6}
L.~Li, M.~Harman, F.~Wu, and Y.~Zhang.
\newblock Sbselector: Search based component selection for budget hardware.
\newblock In {\em International Symposium on Search Based Software
  Engineering}, pages 289--294. Springer, 2015.

\bibitem{lingbo1}
L.~Li, M.~Harman, F.~Wu, and Y.~Zhang.
\newblock The value of exact analysis in requirements selection.
\newblock {\em IEEE Transactions on Software Engineering, PP (99)}, pages 1--1,
  2016.

\bibitem{Manotas:2014:SSE:2568225.2568297}
I.~Manotas, L.~Pollock, and J.~Clause.
\newblock Seeds: A software engineer's energy-optimization decision support
  framework.
\newblock In {\em Proceedings of the 36th International Conference on Software
  Engineering}, ICSE 2014, pages 503--514, New York, NY, USA, 2014. ACM.

\bibitem{bloat2}
N.~Mitchell and G.~Sevitsky.
\newblock The causes of bloat, the limits of health.
\newblock In {\em ACM SIGPLAN Notices}, volume~42, pages 245--260. ACM, 2007.

\bibitem{bloat1}
N.~Mitchell, G.~Sevitsky, and H.~Srinivasan.
\newblock Modeling runtime behavior in framework-based applications.
\newblock In {\em European Conference on Object-Oriented Programming}, pages
  429--451. Springer, 2006.

\bibitem{nagel2017self}
F.~Nagel, G.~M. Bierman, A.~Dragojevic, and S.~Viglas.
\newblock Self-managed collections: Off-heap memory management for scalable
  query-dominated collections.
\newblock In {\em EDBT}, pages 61--71, 2017.

\bibitem{fan2}
J.~Nanavati, F.~Wu, M.~Harman, Y.~Jia, and J.~Krinke.
\newblock Mutation testing of memory-related operators.
\newblock In {\em Eighth {IEEE} International Conference on Software Testing,
  Verification and Validation, {ICST} 2015 Workshops, Graz, Austria, April
  13-17, 2015}, pages 1--10, 2015.

\bibitem{story2}
R.~J. Nowling.
\newblock {Gotchas with Scala Mutable Collections and Large Data Sets}.
\newblock
  \url{http://rnowling.github.io/software/engineering/2015/07/01/gotcha-scala-collections.html},
  2015.
\newblock [Online; accessed 18-February-2017].

\bibitem{petke2017genetic}
J.~Petke, S.~Haraldsson, M.~Harman, D.~White, J.~Woodward, et~al.
\newblock Genetic improvement of software: a comprehensive survey.
\newblock {\em IEEE Transactions on Evolutionary Computation}, 2017.

\bibitem{petke2014using}
J.~Petke, M.~Harman, W.~B. Langdon, and W.~Weimer.
\newblock Using genetic improvement and code transplants to specialise a c++
  program to a problem class.
\newblock In {\em European Conference on Genetic Programming}, pages 137--149.
  Springer, 2014.

\bibitem{poli2008field}
R.~Poli, W.~B. Langdon, N.~F. McPhee, and J.~R. Koza.
\newblock {\em A field guide to genetic programming}.
\newblock Lulu. com, 2008.

\bibitem{Shacham:2009:CAS:1542476.1542522}
O.~Shacham, M.~Vechev, and E.~Yahav.
\newblock Chameleon: Adaptive selection of collections.
\newblock In {\em Proceedings of the 30th ACM SIGPLAN Conference on Programming
  Language Design and Implementation}, PLDI '09, pages 408--418, New York, NY,
  USA, 2009. ACM.

\bibitem{Shacham:2009:CAS:1543135.1542522}
O.~Shacham, M.~Vechev, and E.~Yahav.
\newblock Chameleon: Adaptive selection of collections.
\newblock {\em SIGPLAN Not.}, 44(6):408--418, June 2009.

\bibitem{bloat4}
A.~Shankar, M.~Arnold, and R.~Bodik.
\newblock Jolt: lightweight dynamic analysis and removal of object churn.
\newblock {\em ACM Sigplan Notices}, 43(10):127--142, 2008.

\bibitem{designpattern1}
J.~Vlissides, R.~Helm, R.~Johnson, and E.~Gamma.
\newblock Design patterns: Elements of reusable object-oriented software.
\newblock {\em Reading: Addison-Wesley}, 49(120):11, 1995.

\bibitem{fan3}
F.~Wu, J.~Nanavati, M.~Harman, Y.~Jia, and J.~Krinke.
\newblock Memory mutation testing.
\newblock {\em Information {\&} Software Technology}, 81:97--111, 2017.

\bibitem{fan2015}
F.~Wu, W.~Weimer, M.~Harman, Y.~Jia, and J.~Krinke.
\newblock Deep parameter optimisation.
\newblock In {\em Proceedings of the 2015 Annual Conference on Genetic and
  Evolutionary Computation}, pages 1375--1382. ACM, 2015.

\bibitem{xu2009go}
G.~Xu, M.~Arnold, N.~Mitchell, A.~Rountev, and G.~Sevitsky.
\newblock Go with the flow: profiling copies to find runtime bloat.
\newblock {\em ACM Sigplan Notices}, 44(6):419--430, 2009.

\bibitem{bloat5}
G.~Xu, N.~Mitchell, M.~Arnold, A.~Rountev, E.~Schonberg, and G.~Sevitsky.
\newblock Finding low-utility data structures.
\newblock {\em ACM Sigplan Notices}, 45(6):174--186, 2010.

\bibitem{xu2008precise}
G.~Xu and A.~Rountev.
\newblock Precise memory leak detection for java software using container
  profiling.
\newblock In {\em Software Engineering, 2008. ICSE'08. ACM/IEEE 30th
  International Conference On}, pages 151--160. IEEE, 2008.

\bibitem{1197687}
E.~Zitzler, L.~Thiele, M.~Laumanns, C.~M. Fonseca, and V.~G. da~Fonseca.
\newblock Performance assessment of multiobjective optimizers: an analysis and
  review.
\newblock {\em IEEE Transactions on Evolutionary Computation}, 7(2):117--132,
  April 2003.

\end{thebibliography}

\end{document}